\PassOptionsToPackage{table,dvipsnames}{xcolor}
\documentclass[a4paper,onecolumn, 11pt,unpublished]{quantumarticle}
\pdfoutput=1

\usepackage[utf8]{inputenc}
\usepackage[english]{babel}
\usepackage{amsmath}
\usepackage{hyperref}
\usepackage{qcircuit}
\usepackage{tikz}
\usepackage{lipsum}
\usepackage{multirow}
\usepackage{graphicx}
\usepackage{caption}
\usepackage{stmaryrd}
\usepackage{braket}
\usepackage{amsfonts}       
\usepackage[table]{xcolor} 
\usepackage{subfigure}
\usepackage{amsthm}
\newtheorem{theorem}{Theorem}
\usepackage{enumitem}
\usepackage{algorithmic}
\usepackage[linesnumbered,ruled,vlined]{algorithm2e}%
\usepackage{booktabs}
\usepackage{hyperref}

\newtheorem{assumption}{Assumption}
\newtheorem{conjecture}{Conjecture}

\definecolor{lightgreen}{RGB}{200,255,200}
\definecolor{lightred}{RGB}{255,204,204}

\begin{document}

\title{Learning Neural Decoding with Parallelism and Self-Coordination for Quantum Error Correction}

\author{Kai Zhang}
\orcid{0009-0005-6803-7533}
\affiliation{Department of Computer Science and Technology, Tsinghua University, Beijing, China}
\author{Situ Wang}
\orcid{0009-0009-3012-2248}
\affiliation{Department of Computer Science and Technology, Tsinghua University, Beijing, China}
\author{Linghang Kong}
\affiliation{Zhongguancun Laboratory, Beijing, China}
\orcid{0000-0002-5854-5340}
\author{Fang Zhang}
\affiliation{Zhongguancun Laboratory, Beijing, China}
\orcid{0000-0002-0000-7101}
\email{fangzhang@iqubit.org}
\author{Zhengfeng Ji}
\affiliation{Department of Computer Science and Technology,  Tsinghua University, Beijing, China}
\affiliation{Zhongguancun Laboratory, Beijing, China}
\email{jizhengfeng@tsinghua.edu.cn}
\orcid{0000-0002-7659-3178}
\author{Jianxin Chen}
\affiliation{Department of Computer Science and Technology, Tsinghua University, Beijing, China}
\email{chenjianxin@tsinghua.edu.cn}
\orcid{0000-0002-9365-776X}
\maketitle

\begin{abstract}
  Fast, reliable decoders are pivotal components for enabling fault-tolerant quantum computation. Neural network decoders like AlphaQubit have demonstrated significant potential, achieving higher accuracy than traditional human-designed decoding algorithms. However, existing implementations of neural network decoders lack the parallelism required to decode the syndrome stream generated by a superconducting logical qubit in real time. Moreover, integrating AlphaQubit with sliding window-based parallel decoding schemes presents non-trivial challenges: AlphaQubit is trained solely to output a single bit corresponding to the global logical correction for an entire memory experiment, rather than local physical corrections that can be easily integrated.
  
  We address this issue by training a recurrent, transformer-based neural network specifically tailored for sliding-window decoding. While our network still outputs a single bit per window, we derive training labels from a consistent set of local corrections and train on various types of decoding windows simultaneously. This approach enables the network to self-coordinate across neighboring windows, facilitating high-accuracy parallel decoding of arbitrarily long memory experiments. As a result, we resolve the throughput limitation that previously prohibited the application of AlphaQubit-type decoders in fault-tolerant quantum computation.
  
\end{abstract}

\section{Introduction}

Inspired by Feynman's early vision, the theoretical advantages of quantum computing were solidified three decades ago---evidenced by the provable quantum advantage of Shor's algorithm for factoring~\cite{shor1999polynomial}. Specifically, a system comprising several hundred perfect qubits could efficiently factor the $2048$-bit integers underlying RSA encryption~\cite{gidney2025factor}: a computational task that has long been believed to be intractable for classical computers.
However, across all physical implementations, whether superconducting~\cite{krantz2019quantum}, trapped-ion~\cite{bruzewicz2019trapped}, or other architectures~\cite{bluvstein2024logical, slussarenko2019photonic}, qubits exhibit extreme susceptibility to decoherence induced by environmental perturbations.

To enable reliable quantum computation, quantum error correction codes (QECC) and fault-tolerant quantum computing have been developed~\cite{gottesman1997stabilizer}. These code-based techniques use redundant physical qubits to encode logical qubits, forming a quantitative strategy to overcome the qualitative limitations of qubit fidelity. By suppressing inherent noise and instability to negligible levels, this ``redundancy-by-design'' approach dramatically improves computational accuracy. Although it substantially increases hardware complexity by requiring two to three orders of magnitude more qubits, the fundamental value proposition remains: The benefits brought by the intrinsic speedup of quantum algorithms for problems like factoring can easily outweigh the overhead costs introduced by QECC.

\begin{figure}
    \centering
    \includegraphics[width=0.4\linewidth]{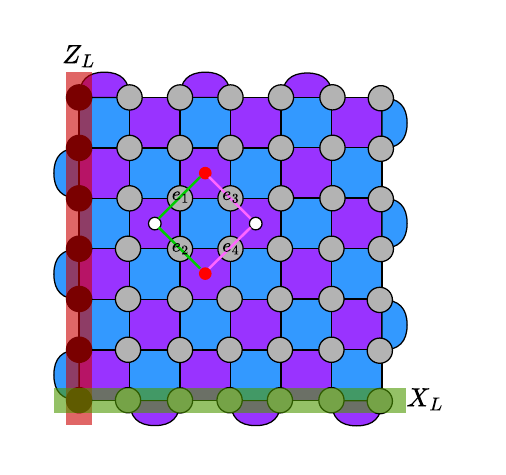}
    \caption{Surface code example for $d = 7$. Data qubits are represented by gray circles. Z stabilizers and X stabilizers are represented by the purple and blue squares (or semi-circle), respectively. Logical Z/X operators are illustrated as red/green rectangles. We also plot red vertices to denote the Z stabilizer syndrome defects, and $\{e_1, e_2, e_3, e_4\}$ to denote 4 possible error chains.}
    \label{fig:surface_code}
\end{figure}

Among all coding schemes that rely on planar connectivity, the surface code~\cite{Dennis_2002} stands out as one of the most promising approaches, as it boasts the highest fault-tolerant threshold~\cite{aharonov1997fault, kitaev1997quantum, Knill1998ResilientQC} currently known. As illustrated in Figure~\ref{fig:surface_code}, a surface code patch typically encodes one logical qubit into a $d \times d$ lattice of physical qubits. The code distance, denoted as $d$, is defined as the weight of the shortest logical operator (red line for the logical Z operator $Z_L$ and green line for the logical X operator $X_L$). As long as the physical error rate $p$ is below the fault-tolerant threshold, the logical error rate can be suppressed exponentially as the code distance increases, thereby providing the desired accuracy for logical quantum computation.

The fault-tolerant threshold depends not only on the code structure and the noise model but also on the decoder. As a purely classical component, the decoder is tasked with outputting a result for each logical measurement in the circuit. To do so effectively, it must continuously monitor the \emph{error syndromes} generated at every step of the computation and track likely error configurations. \emph{Optimal decoding} corresponds to maximum-likelihood decoding (MLD)~\cite{Dennis_2002}, which, in theory, exhaustively evaluates all possible error configurations and selects the logical measurement result with the highest total probability. However, decoding the surface code near-optimally is a surprisingly difficult problem. Minimum weight perfect matching (MWPM)~\cite{Edmonds1973MatchingET, Kolmogorov2009BlossomVA} is a proposed heuristic solution that identifies a single most likely error configuration consisting only of X-type and Z-type errors. While MWPM is ``good enough'' in the sense that it has a provable threshold, there are two fundamental reasons that prevent it from being optimal:
\begin{itemize}
  \item It is a most-likely-error decoder rather than a maximum-likelihood decoder; that is, it identifies a single most likely error configuration instead of the logical equivalence class of error configurations with the highest total probability. In other words, it ignores \emph{degeneracy}—the existence of multiple equivalent error configurations with equal or similar probabilities.
  \item It operates under the principle of decomposing all Pauli errors into X-type and Z-type errors (e.g., a YZ error on a two-qubit gate can be decomposed into an XI error and a ZZ error) and then assumes that all post-decomposition errors are independent random events. In other words, it ignores the \emph{correlation} between post-decomposition errors.
\end{itemize}

The performance gap between MWPM and optimal decoding has been explicitly demonstrated for the repetition code, a low-dimensional analogue of the surface code that admits a polynomial-time maximum-likelihood decoder~\cite{cao2025exact}. For the 2D surface code, truly optimal decoding remains computationally prohibitive except for the smallest code distances and number of rounds, yet existing high-accuracy decoders already confirm that MWPM is far from optimal. As early as 2013, Fowler~\cite{fowler2013optimal} proposed a simple modification to MWPM that incorporates correlation via reweighting. Although Fowler's correlated matching decoder yields only a modest improvement in threshold, it significantly enhances sub-threshold scaling—nearly doubling the benefit of increasing the distance from $d = 3$ to $d = 5$ at $p = 2 \times 10^{-4}$. Higgott~\cite{higgott2023improved} introduced belief-matching, which uses belief propagation for reweighting and improved the fault-tolerant threshold from $0.82\%$ to $0.94\%$ under their error model. Google Quantum AI~\cite{Acharya2022SuppressingQE} developed a tensor network decoder that significantly outperformed belief-matching on both experimental and simulated data, though at a cost of being many orders of magnitude slower. Shutty et al.~\cite{shutty2024efficient} designed Harmony, an ensembled correlated matching decoder that approaches the accuracy of tensor network decoders while being substantially faster and embarrassingly parallelizable.

When it comes to decoding real experimental data as opposed to simulated data generated from an error model, an additional challenge arises: Real quantum hardware contains many qubits and couplers with varying error rates, along with more subtle error sources such as crosstalk and leakage. These error processes are not fully understood~\cite{sivak2024optimization}, and they may not be perfectly captured by a simple hypergraph model. Although traditional decoding algorithms can often succeed with only an approximate error model despite these complications, a machine learning approach is, by definition, the only way to adapt to unknown factors and fully harness the error-correcting potential of the code. Numerous efforts have been made to design neural network decoders for surface codes~\cite{Ni2018NeuralND, liu2019neural, breuckmann2018scalable, Meinerz2021ScalableND, zhang2023scalable, hu2025efficient, cao2025generative}, with AlphaQubit~\cite{bausch2024learning} emerging as a leading example. It employs a recurrent transformer-based architecture and is trained to predict the global logical correction in memory experiments, allowing for fine-tuning with real hardware data. AlphaQubit achieves state-of-the-art performance on Sycamore experimental data for $d = 3$ and $d = 5$, and outperforms correlated matching on simulated data up to $d = 11$.

One disadvantage shared by all known decoders more accurate than MWPM is higher computational cost: In general, there seems to be a tradeoff between a decoder's speed and its accuracy. This is a serious problem because fault-tolerant quantum computation requires real-time decoding~\cite{Zhang2023ACA}. Fortunately, many decoders can be accelerated through sliding-window parallelization~\cite{Tan2022ScalableSD, Skoric2022ParallelWD, bombin2023modular}, which increases the total computational cost by a constant factor, but allows distributing it onto an arbitrary number of decoding units. Unfortunately, this parallelization scheme is not directly applicable to AlphaQubit, because the former makes use of local physical error predictions in order to merge decoding results from adjacent windows, and the latter only returns global logical predictions.

In this work, we aim to train a neural network decoder that retains AlphaQubit's accuracy level while adopting the sliding-window parallel decoding scheme. Inspired by the observation that the merge step is often unnecessary when using the parallel decoding scheme with the MWPM decoder, we train our neural network to output a logical correction bit for each window that can be directly combined without local merging. Even though there is no single ``correct'' way to decode the core region of each individual window, as long as syndromes near the seam are handled consistently by the decoder in adjacent windows, the combined result will be correct. This is evidenced by the fact that the overall logical error rate our decoder achieves for a memory experiment is much lower than the ``error rate'' for each individual window.

The remainder of this paper is structured as follows. Section~\ref{sec:background} introduces the requirement of \emph{real-time decoding}, motivating the sliding-window decoding scheme, and then discusses the rationale for potentially eliminating its merge step. Section~\ref{sec:decoding_scheme} provides a high-level overview of our decoding scheme and illustrates its interaction with the neural network model. Details of the neural network's modules are presented in Section~\ref{sec:model_design}. Section~\ref{sec:model_training} outlines the model's training procedure. Experimental evaluation and analysis of the decoding scheme's performance are covered in Section~\ref{sec:evaluation}. Finally, Section~\ref{sec:discussion} explores the implications and possible extensions of this work.

\section{Background and Motivation}\label{sec:background}

\subsection{Real-time decoding}
A significant fraction of errors in quantum computers arise from decoherence, which happens regardless of whether the qubits are undergoing gates or idling. Consequently, when implementing fault-tolerant quantum computation, the quantum computer cannot just pause and wait for the decoding result: Without periodic syndrome extraction rounds to help locating and isolating decoherence errors, they would accumulate on the data qubits and quickly go over the threshold of the fault-tolerant protocol. In particular, a superconducting quantum computer can execute a round of syndrome extraction every $\sim 1 \mu s$~\cite{RyanAnderson2021RealizationOR}, and considering that the currently demonstrated noise level is barely below threshold~\cite{google2025quantum}, we really do not want to go any slower.

Fortunately, the quantum computer rarely \emph{needs} to wait for the decoding result. Thanks to the Pauli frame technique~\cite{Riesebos2017PauliFF, chamberland2018fault, knill2005quantum}, there is no need to correct every physical error individually and immediately. Instead, the corrections can be tracked by the decoding unit, and applied only to the results of logical measurements~\cite{Zhang2023ACA}. Therefore, the decoder can only stall the quantum computer when the result of an earlier logical measurement is needed to determine what operation needs to be applied now, which happens, for example, when implementing non-Clifford gates with gate teleportation~\cite{litinski2019game}. This problem can be further mitigated by trying to schedule logical operations so that the quantum computer has other tasks to perform while waiting for the result. In the worst case, the quantum computer can stall \emph{on the logical level}, not performing any logical operations but continuing to run the syndrome extraction rounds.

\begin{figure}
  \centering
  \includegraphics[width=0.4\linewidth]{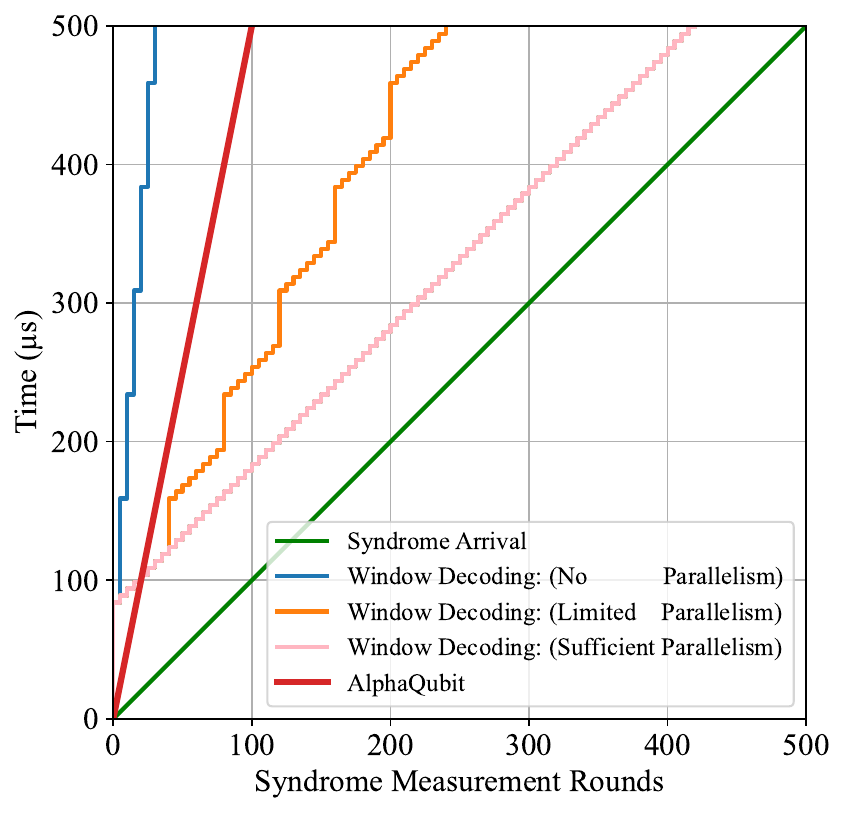}
  \caption{An illustration of the relationship between decoder throughput and latency. At a given round, the latency is the time difference between the decoder response curves and the dark green line representing syndrome generation by the quantum hardware. The latency of AlphaQubit (red line) grows linearly with the number of rounds. For parallel decoders with insufficient throughput (blue and orange lines), the latency also grows nearly linearly and becomes unacceptable. In contrast, for a decoder with sufficient throughput (pink line), the latency remains asymptotically constant regardless of the number of measurement rounds, providing constant feedback latency for logical quantum computation.}
\label{fig:realtime_decoding}
\end{figure}

Still, there is a minimum requirement on the speed of the decoder. If the average speed at which the decoder can process syndromes, the decoding \emph{throughput}, is slower than the rate at which the quantum computer generates them, the decoder will fall progressively further behind. This is illustrated by the red and orange lines in Figure~\ref{fig:realtime_decoding}. The problem is exacerbated by the fact that, as mentioned earlier, the quantum computer continues to generate syndromes even while stalled. Moreover, syndromes generated during a stall for one logical measurement may need to be decoded to determine the \emph{next} logical measurement. If the stall duration increases linearly with the total number of rounds so far, the number of rounds per logical measurement in a dependency chain can grow exponentially~\cite{Terhal2013QuantumEC}, potentially negating any quantum advantage over classical computation. Unfortunately, AlphaQubit~\cite{bausch2024learning} operates in this regime: Although its recurrent architecture allows it to process syndromes as soon as they become available, each round still takes $20 \mu s$ even at $d = 3$, too slow for a superconducting quantum computer operating at $1 \mu s$ per round. AlphaQubit is only trained for memory experiments and thus cannot decode intermediate logical measurements. Even if modified to do so, it would suffer from exponential stalling.

Assume that the single-round inference cost of AlphaQubit can be optimized through techniques such as quantization~\cite{jacob2018quantization} and pruning~\cite{han2015learning}---for instance, a simple FP32 $\to$ INT8 quantization, which can roughly yield a $4\times$ speedup theoretically.
However, Figure~\ref{fig:realtime_decoding} indicates that such an optimization is still insufficient, as the latency grows linearly with the number of syndrome measurement rounds and becomes unacceptable unless the single-round inference latency can be reduced below 1 $\mu s$. This is not scalable in practice, since achieving $\leq 1\mu s$ inference is generally infeasible as $d$ increases. In contrast, if a \emph{parallelized} AlphaQubit implementation were available, where inference can be executed in parallel, the streaming inference latency could be maintained at a constant level. This constant can be further reduced by optimizing the neural network overhead, thus improving the logical feedback speed and the fidelity in lattice surgery~\cite{litinski2019game}.

In other words, a decoder with sufficient throughput does not necessarily eliminate stalling, but keeps it under control. There will still be a \emph{latency} between the last input syndromes and the final decoding results, but over a long computation session, this latency remains asymptotically constant as illustrated in Figure~\ref{fig:realtime_decoding} by the pink line. This means that for every logical measurement, the decoder will at most stall the computation for a constant amount of time, and the overall time cost will remain linear in the size of the quantum circuit.
A prime example of this type of decoder is the parallel sliding-window decoder~\cite{Tan2022ScalableSD, Skoric2022ParallelWD, bombin2023modular}. This decoder necessarily incurs a large initial latency, both to collect enough syndromes to fill a window and to decode them. While parallelization between windows cannot reduce this initial latency, sufficient parallelism---i.e., sufficient throughput---ensures that all subsequent windows are processed with effectively the same latency.

\subsection{Parallel sliding-window decoder without local merging}\label{sec:without_merging}

Existing implementations of the parallel sliding-window decoder~\cite{Tan2022ScalableSD, Skoric2022ParallelWD, bombin2023modular} are usually framed in a hypergraph-based model of decoding. In such a model, a decoding task is represented by a hypergraph $G = (V, E)$, where each edge $e \in E$ corresponds to a physical error source, and each vertex $v \in V$ corresponds to a \emph{detector}, a bit derived from measurement results that indicates the presence of nearby errors. The value of a detector is the \emph{parity} of the number of errors associated with it that actually occurred. In other words, each error \emph{flips} all associated detectors. Given a set of edges $C \subseteq E$, we denote the set of vertices flipped by edges in $C$ as $\partial C$.

A notable characteristic of quantum codes is that \emph{degeneracy} is common: There will be many short \emph{cycles}, i.e., $C \subseteq E$ such that $\partial C = \emptyset$, which represent low-weight \emph{undetectable errors}. However, these undetectable errors do not undermine the fault tolerance of the surface code because they also act trivially on the encoded logical qubit. Concretely, we can define a set of edges $L \subseteq E$ that represents a logical operator, such that a set of errors $\mathcal{E} \subseteq E$ flips this logical operator if and only if $|\mathcal{E} \cap L| \equiv 1 \pmod{2}$. An undetectable error that flips a logical operator must have weight at least $d$, where $d$ is the \emph{fault tolerance distance} of the decoding task.

\begin{figure}
  \centering
  \includegraphics[width=0.7\linewidth]{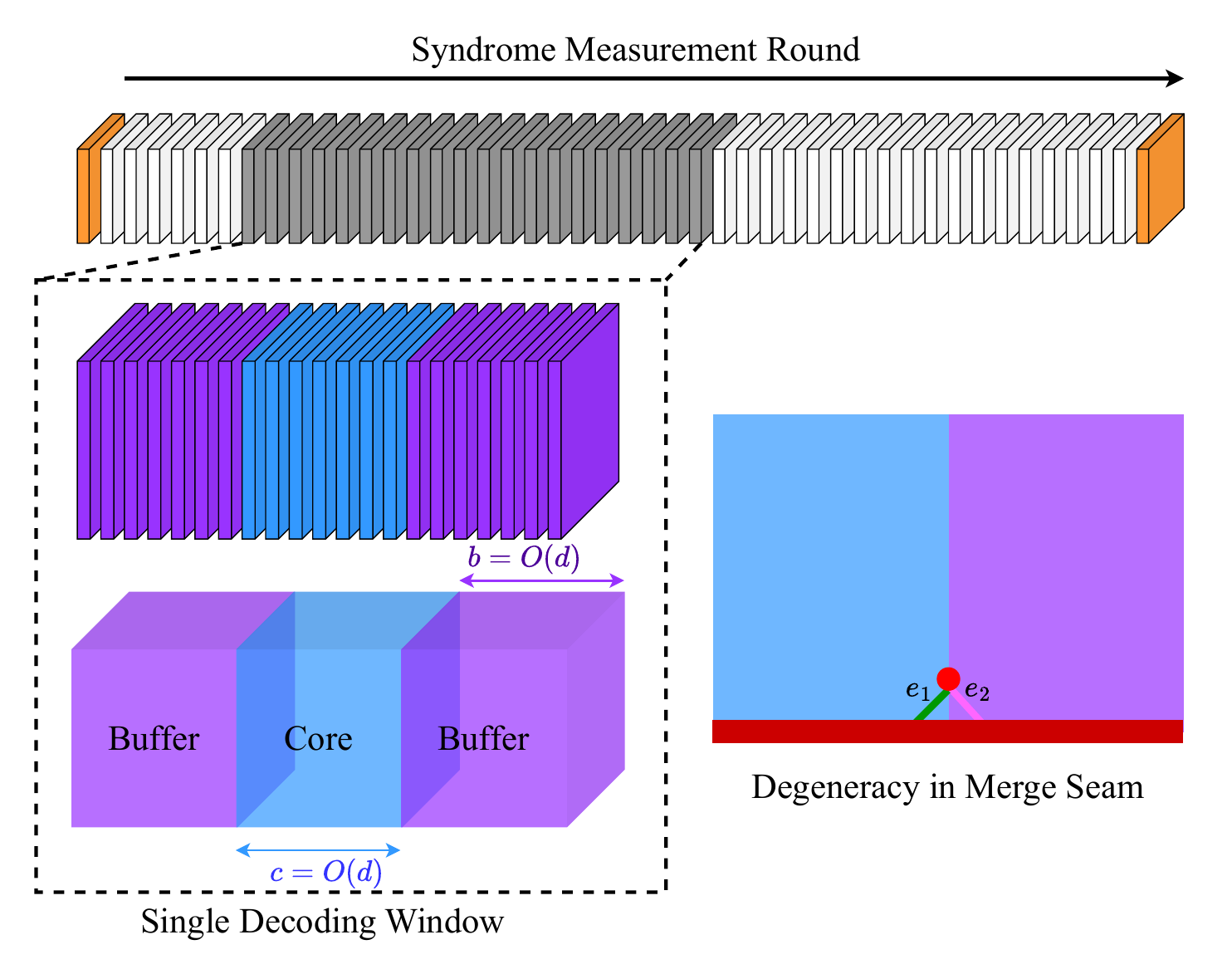}
  \caption{Decoding window visualization for $d = 7$ and degeneracy in merge seam with logical error.}\label{fig:decoding_window}
\end{figure}

One consequence of the ubiquity of degeneracy is that there often exist many equivalent ways to correct the same error syndrome. Figure~\ref{fig:surface_code} shows an example: Let $C = \{e_1, e_2, e_3, e_4\}$ be a length-4 cycle such that $|C \cap L| \equiv 0 \pmod{2}$, and $C_1 = \{e_1, e_2\}$ and $C_2 = \{e_3, e_4\}$ be two error configurations, then $\partial C_1 = \partial C_2$ and $|C_1 \cap L| \equiv |C_2 \cap L| \pmod{2}$, meaning that the same set of detector flips can be corrected either as $C_1$ or as $C_2$, but that is fine because both corrections have the same effect on the logical operator.

The basic idea of sliding-window decoders is to break the decoding graph $G$ into windows, such that each decoding process only makes the decision on a subset of edges, called the \emph{core region} of a window. To ensure the accuracy of decoding results, each window contains not only the detector information within the core region, but also a \emph{buffer region} of thickness $O(d)$ around the core region, as shown in Figure~\ref{fig:decoding_window}.

Degeneracy can become a problem because adjacent windows may output two sets of corrections that are globally equivalent, but not equivalent when the boundary of the core region is considered. We illustrate this issue also in Figure~\ref{fig:decoding_window}: Suppose $C = \{e_1, e_2\}$ and let $C \cap L = \{e_1, e_2\}$, $e_1$ is in the core region of window 1 and $e_2$ is in the core region of window 2. If window 1 outputs the correction $C_1 = \{e_1\}$ and window 2 outputs the correction $C_2 = \{e_2\}$, then both $e_1$ and $e_2$ will be corrected in their respective window's core regions and the logical operator will be flipped twice---end up not correcting the logical observable. This will result in a logical error if the actual error configuration is $C_1$ or $C_2$, since the ground truth logical operator should be flipped only once.

Several schemes exist for addressing this issue.
Ref.~\cite{Tan2022ScalableSD} handles this degeneracy problem by \emph{merging}
the corrections at the \emph{seam} between core regions.
This is achieved by combining the local corrections (i.e., selected edges)
within each window's core region, identifying detectors that remain flipped
after corrections, and decoding them again on a 2D ``seam decoding graph''.
Ref.~\cite{Skoric2022ParallelWD} also employs two rounds of decoding; although
the seam decoding graphs are extended into 3D windows with fixed boundary
conditions, the underlying principle remains the same.
Ref.~\cite{bombin2023modular} further shortens the core regions in the first
round of windows, so that the primary purpose of the initial decoding round
becomes determining an appropriate set of boundary conditions for the second
round.

Despite these differences in detail, all these schemes require the base decoder to output a set of local corrections and to enforce consistency between the outputs of adjacent windows. This approach is incompatible with the design of the state-of-the-art neural network decoder, AlphaQubit, which does not rely on explicit knowledge of the decoding graph and outputs only a single bit, the global logical correction.

\begin{figure}
\centering
\includegraphics[width=0.8\linewidth]{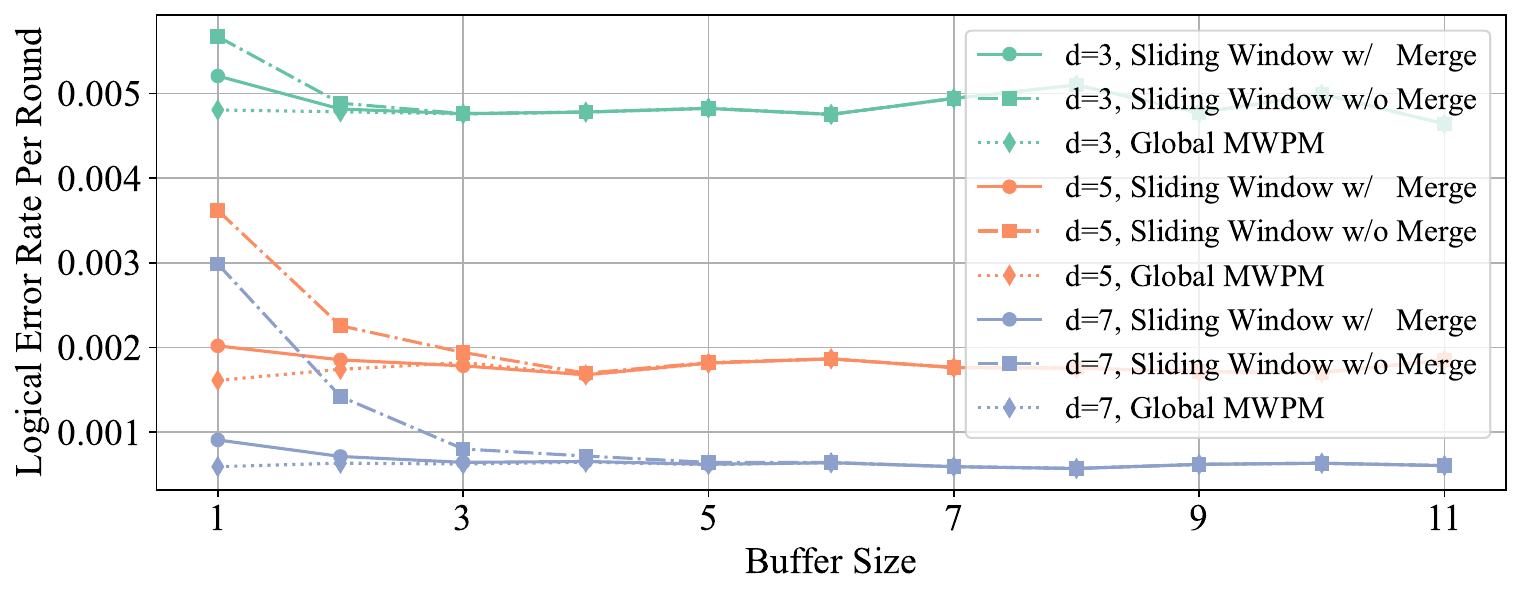}
\caption{Parallel window decoding without merge. Here we conduct memory experiments up to about $100$ syndrome measurement rounds under $p=0.003$, each data point with the exactly same random seed to ensure fairness. The decoding accuracy of sliding window decoders (MWPM as the inner decoder) with or without the merge operation converges with the global MWPM decoding as the buffer size approaches the code distance $d$.}
\label{fig:decode_without_merge}
\end{figure}

Our key observation is that, when using the scheme of~\cite{Tan2022ScalableSD} with the MWPM decoder, as long as the length of the buffer region is sufficiently large (approximately $d$ rounds), the probability of any remaining flipped detectors requiring merging becomes extremely low compared to the logical error rate. It turns out that the MWPM decoder typically does not decode the same set of error syndromes differently, so merges occur only with complex error configurations spanning the entire length of the buffer (see Appendix~\ref{sec:no_merge} for a detailed discussion). Figure~\ref{fig:decode_without_merge} also provides experimental confirmation of this observation.

This observation inspires us to train a neural network decoder whose output can be used in the sliding-window decoding scheme without merging. Similar to AlphaQubit, our decoder outputs only a single bit of logical correction per window; however, as long as these bits are based on the neural network's internal decisions, which are consistent across windows, they can be simply XORed together to obtain a global correction for the entire experiment without issues.

In the following sections, we describe our model, including detailed explanations of its architecture, a novel supervision framework tailored to parallel window decoding, and the corresponding training methodology.

\section{Decoding Scheme Overview}\label{sec:decoding_scheme}

\begin{figure}
\centering
\includegraphics[width=\linewidth]{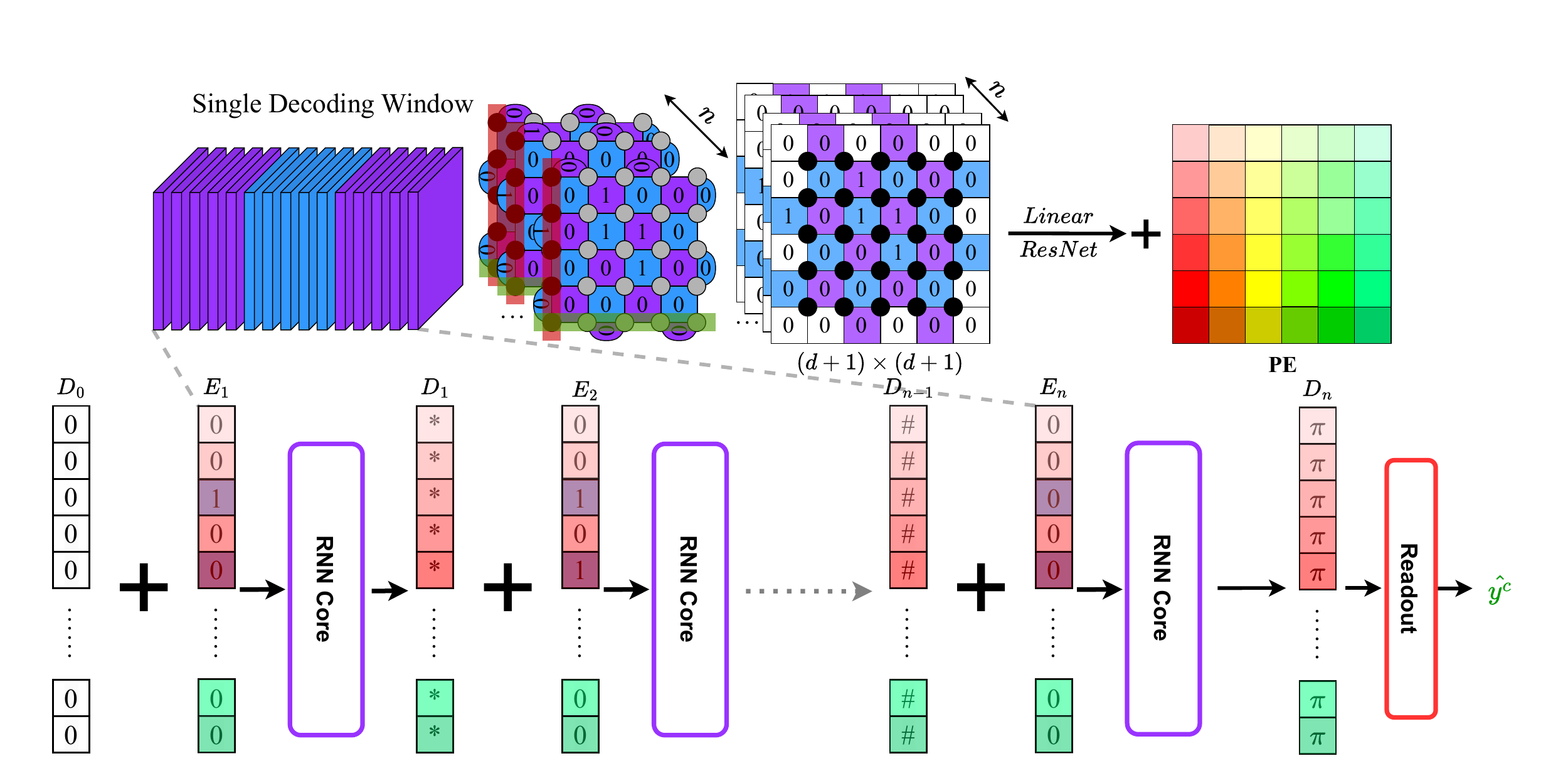}
\caption{Model pipeline. Here we illustrate the model pipeline with $d=5$ syndrome as input. Each decoding window containing $n = b + c + b$ syndrome measurement rounds will be embedded and go through the \emph{RNN Core} for $n$ times, quite similar to AlphaQubit. However, the final output logit is designed to predict the logical flip only in the core region of the decoding window.}
\label{fig:model_pipeline}
\end{figure}

In the parallel decoding scheme, the input syndrome stream from the quantum hardware is cut into overlapping sliding windows. Our neural network model will take syndromes from one decoding window as input, and output a single bit representing the contribution of the core region of this window to the overall logical correction. Although the first and last decoding windows in a memory experiment are slightly different, we pad them appropriately so that they can be handled by the same network (see Section~\ref{sec:padding} for details). Figure~\ref{fig:model_pipeline} is an illustration of the overall decoding scheme.

In order to perform supervised training and enable the model to predict logical errors only in the core region of a decoding window, we use a ``local ground truth'' $\mathcal{E}$, the set of edges in the decoding graph which have been flipped (i.e., which \emph{physical} errors have happened). Even though this information cannot be observed in real experiments, it is available when the training data is generated through simulation. The label for each decoding window in the training data is then derived as:
\begin{equation}\label{eq:core_logical_error}
  y_i = |\mathcal{E} \cap E^c_i \cap L| \bmod 2
\end{equation}
where $E^c_i$ is the set of edges in the core region of window $i$, and $L$ is a global logical operator representative. Since $\{E^c_i\}$ is a partition of the original $E$, it is guaranteed that:
\begin{equation}
    \bigoplus_{i=1}^{n} y_i = y = |\mathcal{E} \cap L| \bmod 2.
\end{equation}
In other words, if the neural network correctly predicts $y_i$ for each window $i$, then combining all predictions with XOR yields the correct overall logical correction for the memory experiment. Note that the converse is not necessarily true: If the network flips an even number of $y_i$, the overall prediction $y$ would still be correct. However, as mentioned in Figure~\ref{fig:decoding_window} before, degeneracy may cause independently trained neural networks to fail, since the same defect can correspond to different logical labels of the core region, making it difficult for the neural network to learn and converge. This represents one of the key challenges that we must overcome in the following sections.

\section{Model Design}\label{sec:model_design}

\begin{figure}
\centering
\includegraphics[width=\linewidth]{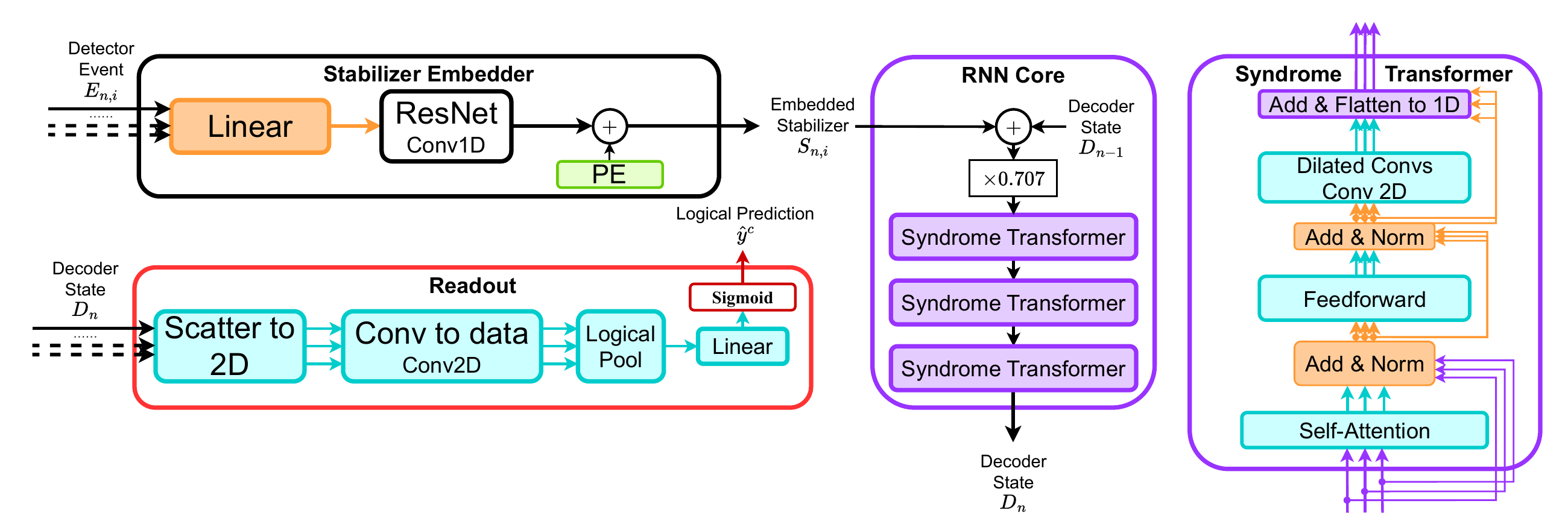}
\caption{Neural network model architecture in detail.}
\label{fig:neural_network_architecture}
\end{figure}

\subsection{Model overview}

We follow AlphaQubit's model design but simplify it with minor modifications, as shown in Figure~\ref{fig:neural_network_architecture} and Table~\ref{tab:model_parameter}:
\begin{enumerate}
  \item Our model only takes hard detector events as input at this moment for
        simplicity.
        However, adding measurement, soft and leakage
        information~\cite{Acharya2022SuppressingQE, varbanov2025neural,
        bausch2024learning} is available and may improve performance in the
        future.
  \item The positional encoding (\textbf{PE}) is applied after the \emph{ResNet}
        module, instead of before.
  \item The readout module has one less \emph{ResNet} module compared to
        AlphaQubit.
  \item We directly utilize the \emph{Feedforward} layer the same
        as~\cite{vaswani2017attention} rather than the ``dense block augmented
        with gating''~\cite{bausch2024learning} architecture used in AlphaQubit.
\end{enumerate}
In the following sections, we will detail the subtle differences from the original model design in AlphaQubit and the corresponding considerations.

\begin{table}
\centering
\caption{Module parameters setting.}
\small
\label{tab:model_parameter}
\begin{tabular}{c c c}
\toprule
\textbf{Module} & \textbf{Layer} & \textbf{Parameters} \\
\midrule
\multirow{2}{*}{\textbf{StabilizerEmbedder}} 
    & Linear & $\text{Linear}(1 \rightarrow d_{\text{model}})$ \\
    & ResNet 1D Convolution & 2 $\times$ Conv1d($d_{\text{model}}, d_{\text{model}}, k=3$) \\
\midrule
\multirow{3}{*}{\textbf{SyndromeTransformer}} 
    & Multi-Head Self Attention & $d_{\text{model}}, n_{\text{head}}$ heads \\
    & Feedforward & $d_{\text{model}} \rightarrow 5d_{\text{model}} \rightarrow d_{\text{model}}$ \\
    & Dilated Convolutions & 3 $\times$ Conv2d($d_{\text{model}}, d_{\text{model}}, k = 3$)\\
\midrule
\multirow{1}{*}{\textbf{RNNCore}} 
    & Transformer Layers & 3 $\times$ \textbf{Syndrome Transformer} \\
\midrule
   \multirow{3}{*}{\textbf{Readout}} & 2D Convolution & 1 $\times$ Conv2d ($d_{\text{model}},  d_{\text{model}}, k=2$)\\
    & Logical Pooling & ($d \times d \to d$) \\
    & Linear & Linear($d \cdot d_{\text{model}}  \rightarrow 1$)\\
\bottomrule
\end{tabular}
\end{table}

\subsection{Syndrome embedding}
\subsubsection{Boundary padding strategy}

\begin{figure}
\begin{center}
\includegraphics[width=0.8\linewidth]{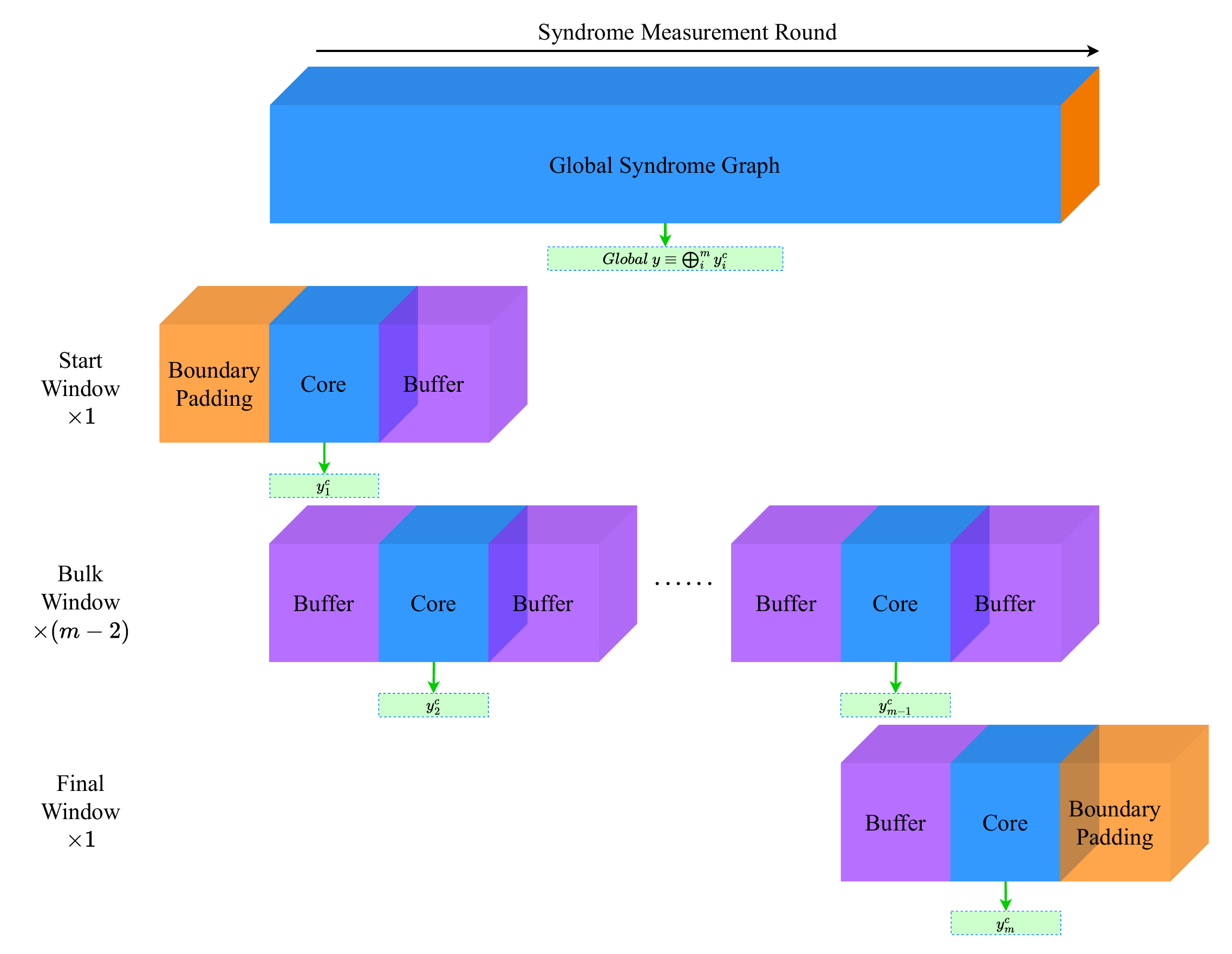}
\caption{Illustration for the global syndrome graph in quantum memory experiment and three kinds of decoding windows. Suppose the global syndrome graph contains $N$ syndrome measurement rounds and can be partitioned into $m$ decoding windows, then the decoding windows can be divided into $1$ \emph{start window}, $1$ \emph{final window} and $m-2$ \emph{bulk window}s.}
\label{fig:masked_window_embedding}
\end{center}
\end{figure}

\label{sec:padding}
In the memory experiment, the first and the final syndrome measurement rounds are ``closed time boundaries'' as described in~\cite{Tan2022ScalableSD}. This means that neither a single $Z$ detector flip in the first round nor one in the final round cannot be interpreted as an ancilla measurement error, since the logical qubit is initialized by preparing all \emph{data} qubits to $|0\rangle$ and measured by measuring all \emph{data} qubits in the $Z$ basis. However, when dividing a long decoding graph into separate decoding windows, intermediate decoding windows will have ``open time boundaries'', and thus require buffer regions for accurate decoding. Therefore, there are three different types of decoding windows as depicted in Figure~\ref{fig:masked_window_embedding}: the \emph{start window}, the \emph{bulk window}s, and the \emph{final window}.

In works such as~\cite{Tan2022ScalableSD}, the total number of syndrome rounds in each window is kept constant (as much as possible). This means that the start window and the final window have longer core regions since they each only have one buffer region instead of two. We instead opt to align the semantics of all three types of windows by keeping the size of the \emph{core regions} constant. For the start window and the final window, we add boundary padding (orange regions in Figure~\ref{fig:masked_window_embedding}) filled with zeros in place of the nonexistent buffer regions. Since a zero-filled padding naturally suppresses matching across the time boundary, we no longer need to define special semantics for those ``closed time boundaries''. This ensures that the three window types have consistent semantics and shapes, simplifying our neural network design while improving load balancing in parallel decoding.

Subsequent experiments demonstrate the effectiveness of this streamlined approach. Instead of applying specialized encoding to the three different window types, we align their semantics in the data, enabling a single network to handle different window types and thereby improving the training efficiency. Moreover, joint learning across the three window types further enhances the self-consistency of the network model.

\subsubsection{Lattice-based syndrome encoding}

To capture the geometric structure of the surface code, we map the locations of all stabilizers in a distance-$d$ surface code into a $(d + 1) \times (d + 1)$ array, as shown in Figure~\ref{fig:model_pipeline}. Each round of syndrome data can then be naturally embedded into a $(d + 1) \times (d + 1)$ tensor $\mathbf{E}$, with $1$ representing a syndrome defect (detector flip) and $0$ for the absence of a defect. For locations where no syndrome vertices exist, we also use 0 as a natural padding value. We prepare our data for use in the transformer architecture by flattening each $(d + 1) \times (d + 1)$ tensor into a vector. An entire decoding window is thus encoded into a sequence of $b+c+b$ vectors, where $b$ is the buffer size and $c$ is the core region size. The temporal aspect of the data will be handled by the RNN component of the model.

\subsubsection{Syndrome convolution}\label{sec:conv1d}

We move our simplified \emph{ResNet} layers, \texttt{Conv1D} with residual connection~\cite{He2015DeepRL}, to the front of the positional encoding layer (\textbf{PE}). This modification is motivated by our practical observations: Since convolutional layers do not inherently require \textbf{PE}, we believe that applying \texttt{Conv1D} directly to the flattened syndrome data can better capture local error features, without the loss of original information that \textbf{PE} might introduce. The output of the \emph{ResNet} is a sequence of tensors $\mathbf{E}_\text{convolved}$ with shape $(d + 1)^2 \times d_\text{model}$.

\subsubsection{Positional encoding}
\label{sec:positional_encoding}

To help the transformer module capture spatial relationships, the encoder augments the convolutional representation with a deterministic positional encoding~\cite{vaswani2017attention}. A fixed sinusoidal encoding \(\mathbf{PE}(t)\in\mathbb{R}^{d_\text{model}}\) is generated as:

\begin{equation}
\mathbf{PE}_{2i}(t) = \sin \Bigl(\dfrac{t}{10000^{2i/d_\text{model}}}\Bigr), \ \mathbf{PE}_{2i+1}(t) =\cos \Bigl(\dfrac{t}{10000^{2i/d_\text{model}}}\Bigr)
\end{equation}
where $i = 0,1,\dots,\left\lfloor d_\text{model}/2\right\rfloor$ and $t = 0,1,\dots,(d+1)^2-1$. This yields a spectrum of $d_{model}$ sinusoid sequences with length $(d+1)^2$, whose wavelengths form a geometric progression. The final stabilizer embedding $\mathbf{S}$ is obtained by the element-wise summation
\begin{equation}
\mathbf{S} = \mathbf{E}_\text{convolved} + \mathbf{PE}.
\end{equation}
This embedding endows the model with translation-equivariant yet position-sensitive features, capturing local error patterns while preserving global location information.

\subsection{RNN core}
The \emph{RNN core} receives the complete decoding window syndrome data after undergoing the embedder, including the left buffer region, the core region, and the right buffer region, one round at a time in temporal order. The initial decoder state $\mathbf{D_0}$ is set to be zero for all kinds of decoding windows. At round $n$, the RNN core combines the input stabilizer embedding tensor  $\mathbf{S}_{n}$ with the current decoder state $\mathbf{D_{n-1}}$, normalized by $\frac{\sqrt{2}}{{2}}$. This combined input is then processed through three sequential \emph{syndrome transformer} layers, and the output becomes the new decoder state:
\begin{equation}
    \mathbf{D_n} = \texttt{RNNCore}((\mathbf{S_{n}} + \mathbf{D_{n-1}}) \times 0.707).
\end{equation}

The \emph{syndrome transformer} module inside the \emph{RNN core} is primarily designed to capture more global error information, with the \emph{self-attention} layer as its core component, thus we just follow the standard implementation in~\cite{vaswani2017attention}. We also follow the implementation in~\cite{bausch2024learning}, where the output after attention is scattered onto a 2D grid for \emph{dilated convolution}, thereby further aggregating error information over the surface code topology. Both the input and the final output of the syndrome transformer module has the same shape as the embedded syndrome input, enabling them to be easily chained together.

\subsection{Readout}
After processing all rounds of a decoding window, the final decoder state $\mathbf{D_{n}} = \mathbf{D_{b+c+b}}$ is passed to the \emph{readout} module. The readout module first shrinks the dimension from $(d + 1) \times (d + 1)$ to $d \times d$ using a \emph{Conv2D} layer without padding. Then, the $d \times d$ grid is pooled only in the direction perpendicular to the logical operator, and the resulting $d \times d_\text{model}$ feature tensor is projected to a singular logit via a simple linear layer as the final logical prediction $\hat{y}^c_i$ for the core region. The entire process, from the syndrome input to the logical readout, is also comprehensively outlined in Algorithm~\ref{alg:rnn_decoder_alg}, providing a clearer and more structured overview.

\begin{algorithm}[H]
\caption{Window Neural Decoding}
\label{alg:rnn_decoder_alg}
\KwIn{$\mathbf{E} \in \mathbb{R}^{B \times (b + c + b) \times (d + 1) \times (d + 1)}$: Input detector events (soft or leakage information are compatible as well)}
\KwOut{$\hat{y}$: Logical predictions without or with recurrent training strategy}
\BlankLine
Initialize $decoder\_state \ \mathbf{D_0} \gets \mathbf{0} \in \mathbb{R}^{B \times (d + 1)^2 \times d_{model}}$\;
\If{not recurrent\_training}{
    \For{$n \gets 1$ \KwTo $b + c + b$}{
        $\mathbf{E_n} \gets \mathbf{E}[:, n]$\;
        $\mathbf{S_n} \gets \textbf{StabilizerEmbedder}(\mathbf{E_n})$\;
        $\mathbf{D_n} \gets \textbf{RNN\_Core}(\mathbf{S_n}, \mathbf{D_{n-1}})$\;
    }
    $\hat{y} \gets \textbf{Readout}(\mathbf{D_n})$\;
    \Return{$\hat{y}$}\;
}
\Else{
    $\hat{y} \gets [\ ]$\;
    \For{$n \gets 1$ \KwTo $b + c + b$}{
        $\mathbf{E_n} \gets \mathbf{E}[:, n]$\;
        $\mathbf{S_n} \gets \textbf{StabilizerEmbedder}(\mathbf{E_n})$\;
        $\mathbf{D_n} \gets \textbf{RNN\_Core}(\mathbf{S_n}, \mathbf{D_{n-1}})$\;
        \If{$n > 2 * b$}{
            $\hat{y}_{single} \gets \textbf{Readout}(\mathbf{D_n})$\;
            Append $\hat{y}_{single}$ to $\hat{y}$\;
        }
    }
    \Return{$\hat{y}$}\;
}
\end{algorithm}

\section{Model Training}
\label{sec:model_training}

\subsection{Singular prediction training}
As the logical error prediction is exactly a binary classification problem, we use the BCE loss for all kinds of decoding windows:

\begin{equation}
Loss = - \frac{1}{B} \sum_{i=1}^{B} \left[y^c_{i} \log(\hat{y}^c_{i}) + (1 - y^c_{i}) \log(1 - \hat{y}^c_{i}) \right]
\end{equation}
where $B$ is the batch size, $y^c_i$ is the ground truth logical error flip in the core region of a decoding window, acquired from the DEM simulator~\cite{Gidney2021StimAF}, and $\hat{y}^c_{i}$ is the window prediction.

\subsection{Multi-layer recurrent training}
\label{sec:recurrent_training}

To facilitate the training process and guide the network to progressively learn the ability of predicting the core region, we design a recurrent training method as in Algorithm~\ref{alg:rnn_decoder_alg}, which essentially allows the network to simultaneously train with different ``truncated'' core region sizes $\tau = 1, 2, \dots, c$. This is especially useful for training larger code distances like $d = 7$, as directly learning $b+c+b$ rounds of syndrome data may be too challenging.

In the recurrent training mode, for each $\tau = 1, 2, \dots, c$, the first $b+\tau+b$ rounds of data in a decoding window are treated as if they were a complete decoding window with $b$ rounds of left buffer, $\tau$ rounds of core region, and $b$ rounds of right buffer. Therefore the middle $\tau$ rounds of ground truth are used to derive a label $y_i^\tau$, and the neural network also predicts a $\hat{y}_i^\tau$. Thanks to the recurrent structure of the RNN core, $\hat{y}_i^\tau$ can be computed by simply truncating the intermediate decoder state $D_{b+\tau+b}$ to the readout module, and thus all $\hat{y}_i^\tau$ can be computed with the same $b+c+b$ invocations of the RNN core, as demonstrated in Algorithm~\ref{alg:rnn_decoder_alg}. The loss formula of a batch becomes

\begin{equation}
Loss = - \frac{1}{Bc} \sum_{i=1}^{B} \sum_{\tau=1}^{c} \left[y^\tau_i \log(\hat{y}^\tau_i) + (1 - y^\tau_i) \log(1 - \hat{y}^\tau_i) \right].
\end{equation}

\subsection{Training process}
Training the neural network decoder, especially for larger code distance and lower physical error rate, demands substantial computational resources. We utilize up to 10 NVIDIA GTX 4090 GPUs to train models for $d = 3, 5, 7$---a configuration sufficient for our proof-of-principle demonstration. All training data are generated randomly using Stim~\cite{Gidney2021StimAF} by two steps: First, we generate the global syndrome data with about $3d\sim5d$ syndrome measurement rounds. Second, we split the global syndrome data into $3\sim5$ decoding windows, and mix them together as our training dataset. In our training process, we generally adopt a larger physical error rate such as $p \sim 0.005$ at the beginning, which enables the model to more quickly learn difficult decoding scenarios and facilitate convergence. Nevertheless, we observe that, in the fine-tuning stage, using a smaller error rate like $p \sim 0.001$ is advantageous.

Here we present the hyperparameters, model size and the approximate training time in terms of GPU hours in Table~\ref{tab:training_gpu_hours}. During our training process, we observed behavior consistent with a previously reported finding~\cite{bausch2023learning}---namely, that the number of training samples needed to match the performance of classical algorithms~\cite{higgott2025sparse, Higgott2021PyMatchingAP} grows exponentially with code distance. We further discovered an additional key trend: As the testing physical error rate $p$ decreases, the required number of training samples also increases exponentially. More detailed training records are provided in Appendix~\ref{sec:train_log}.

\begin{table}[ht]
\centering
\caption{Hyperparameter setting and training overhead for different code distances. Hyperparameters include the feature dimension $d_\text{model}$, the number of attention heads $n_{head}$, the batch size $B$, and the learning rate $\alpha$.}
\label{tab:training_gpu_hours}
\begin{tabular}{|c|c|c|c|c|c|c|c|c|}
\hline
\multirow{2}{*}{$d$} & \multicolumn{6}{c|}{Hyperparameters} & \multirow{2}{*}{Model Size$^*$} & \multirow{2}{*}{GPU Hours (h)} \\
\cline{2-7}
 & $d_{\text{model}}$ & $n_{head}$ & $B_\text{init}$ & $B_\text{final}$ & $\alpha_\text{init}$ & $\alpha_\text{final}$ & & \\
\hline
3 & 192 & 4 & 256 & 512 & 1e-4 & 5e-5 & 4,586,113 & 40 \\
5 & 192 & 4 & 576 & 576  & 1e-4 & 5e-5 & 4,586,497 & 300 \\
7 & 192 & 4 & 448 & 1344  & 5e-4 & 5e-5 & 4,586,881 & 2000 \\
\hline
\end{tabular}
\begin{flushleft}
\footnotesize
$^*$ The difference in model size arises from slight variations in the parameters of the projection network in the \emph{readout} module corresponding to different code distances.
\end{flushleft}
\end{table}

\section{Evaluation}
\label{sec:evaluation}
\subsection{Experimental setup}

\subsubsection{Error model}
\label{sec:error_model}
We use the same circuit-level noise model as in~\cite{Tan2022ScalableSD}, where each reset operation intending to set a qubit to $|0\rangle$ (resp. $|+\rangle$) has probability $p$ to set it to $|1\rangle$ (resp. $|-\rangle$) instead, each measurement has probability $p$ to get the result flipped, and each CNOT gate and idle gate is followed by a depolarizing channel with strength $p$. Idle gates are inserted during CNOT rounds on qubits near the boundary not participating in a CNOT gate, and on all data qubits twice while ancilla qubits are undergoing measurement and then reset.

\begin{figure}[htbp]
\begin{center}
\subfigure[Event density of our noise model]{
    \includegraphics[width=0.4\linewidth]{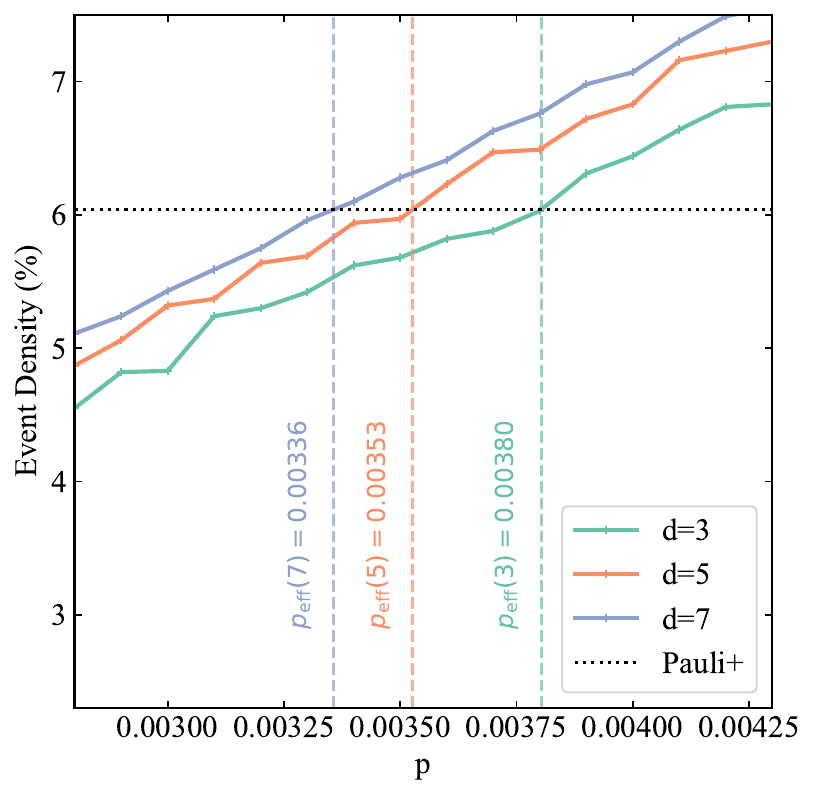}
    \label{fig:event_density}
}
\subfigure[MWPM threshold of our noise model]{
    \includegraphics[width=0.4\linewidth]{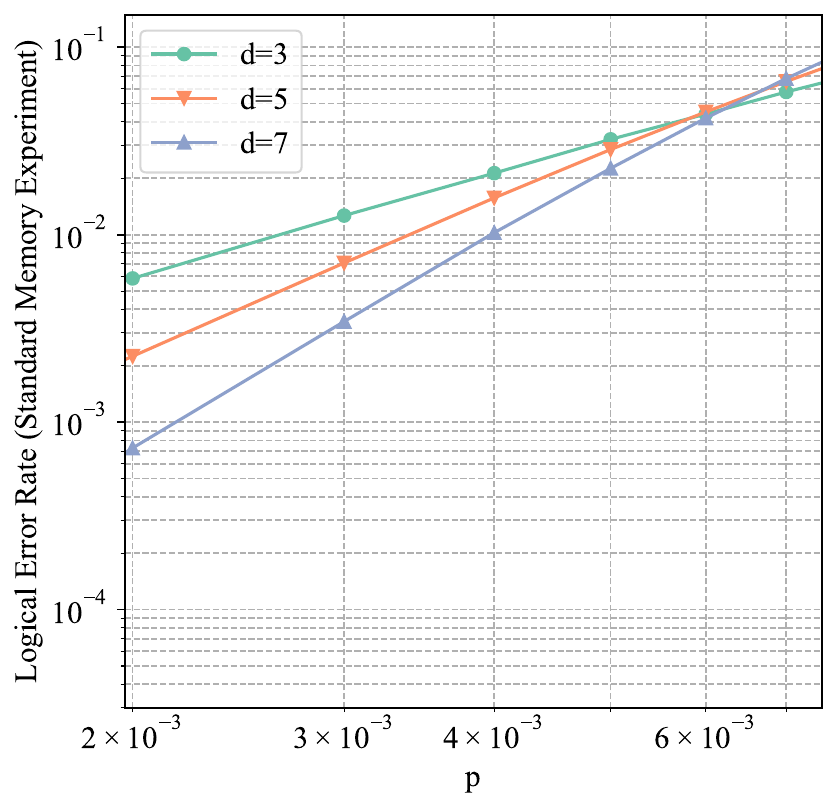}
    \label{fig:threshold}
}
\caption{Event density of our noise model and the corresponding MWPM threhold. (a) The effective physical error rate $p_\text{eff}(d)$ for code distance $d$ is calculated as the physical error rate where the detector events achieve the same density with Pauli+ noise model. Here we get $p_\text{eff}(3), p_\text{eff}(5), p_\text{eff}(7) = \{0.00380, 0.00353, 0.00336\}$, respectively. (b) The corresponding MWPM threshold for our noise model is about $0.6\%$ for a standard surface code memory experiment with $d$ syndrome measurement rounds.}
\end{center}
\end{figure}

In order to evaluate our decoder under roughly the same noise levels as AlphaQubit's Pauli+ dataset~\cite{Acharya2022SuppressingQE, google2025quantum, bausch2024learning, bausch2023learning}, we use the detector event density as a measure of the noise level. We estimate the detector event density values of our error model under different code distance $d$ and physical error rate $p$, and match them to the value reported in~\cite{bausch2024learning}, as shown in Figure~\ref{fig:event_density}. We compute an effective physical error rate $p_\text{eff}$ for each $d$, and use this noise level in our benchmark experiments. All training and test data are generated using Stim~\cite{Gidney2021StimAF}.

\subsubsection{Decoding configuration and comparison targets}
For all our experiments, we set both the buffer region size $b$ and the core region size $c$ to the code distance $d$. Our previous observations (see Figure~\ref{fig:decode_without_merge}) suggest that $b \ge d$ is necessary for sliding-window decoding without merging to work, and then we want $c = \Theta(d)$ to keep the parallelization overhead a constant factor.

We evaluate the decoding accuracy of our neural network decoder 
against global decoding algorithms, including PyMatching~\cite{higgott2025sparse, Higgott2021PyMatchingAP} and Belief-Matching~\cite{higgott2023improved}. PyMatching represents the state-of-the-art pure MWPM decoder, with an exact implementation of MWPM and almost $\mu s$ level decoding latency on modern CPUs. Belief-Matching is a combination of belief-propagation (BP) and MWPM, with higher accuracy than pure MWPM but much lower decoding speed. The software implementation of Belief-Matching~\cite{higgott2023improved} we are using executes up to 20 iterations of BP by default; if BP does not converge, the final state of the BP solver is used to reweight the decoding graph for MWPM. Our preliminary tests show that increasing the iteration cap above 20 has a negligible impact on the accuracy, but further increases the average decoding latency, which is already on the order of milliseconds with 20 iterations. Therefore, we keep this default configuration. Both PyMatching and Belief-Matching utilize DEM corresponding to Section~\ref{sec:error_model} directly from Stim as the decoding prior to ensure fairness.

\subsection{Individual window analysis}
\label{sec:individual_window}
We first study how well the neural network predicts $y_i$ for each window $i$. For $d = 3, 5, 7$, we simulate a minimal memory experiment with $N = 3d$ measurement rounds, which will be divided into exactly one start, one bulk, and one final decoding window by our decoding scheme. We feed all decoding windows to our neural network to get predictions $\hat{y}_i$, and compare them with $y_i$ to get a ``logical error rate'' $p_i = \Pr[\hat{y}_i \ne y_i]$ for each type of windows. We also calculate the global logical error rate $p_g = \Pr[\hat{y}_1 \oplus \hat{y}_2 \oplus \hat{y}_3 \ne y_1 \oplus y_2 \oplus y_3]$ for this minimal memory experiment. The results are presented in Table~\ref{tab:independent_decoding}.

\begin{table}[h]
\centering
\caption{Logical error rates for the individual windows and the entire $N = 3d$ memory experiment. The column labeled $\hat{p}_g$ is the estimated global logical error rate from the error rates of each window, assuming that those ``errors'' happen independently between windows.}
\label{tab:independent_decoding}
\begin{tabular}{|c | c | ccccc|}
\hline
\multirow{3}{*}{$d$} & \multirow{3}{*}{$p$} & \multicolumn{5}{c|}{Logical Error Rate} \\
\cline{3-7}
 &  & $p_1$ & $p_2$ & $p_3$ & $p_g$ & $\hat{p}_g$\\
 &  & Start Window & Bulk Window & Final Window & Global & Estimated\\
\hline
3 & 0.00380 & 0.0203 & 0.0256 & 0.0165 & \cellcolor{lightgreen}0.0434 & \cellcolor{lightred}0.0599\\
5 & 0.00353 & 0.0158 & 0.0244 & 0.0157 & \cellcolor{lightgreen}0.0228 & \cellcolor{lightred}0.0539\\
7 & 0.00336 & 0.0151 & 0.0249 & 0.0144 & \cellcolor{lightgreen}0.0065 & \cellcolor{lightred}0.0525\\
\hline
\end{tabular}

\end{table}

We observe that these results show that these ``mispredictions'' are not independent between windows. If they were, then the global logical error rate should be given by:
\begin{equation}
    \hat{p}_g = p_1 (1-p_2)(1-p_3) + (1-p_1) p_2 (1-p_3) + (1-p_1)(1-p_2) p_3 + p_1 p_2 p_3.
    \label{eq:independent_decoding}
\end{equation}
In Table~\ref{tab:independent_decoding}, we see that $p_g$ (green shaded cells) is consistently lower than $\hat{p}_g$ (red shaded cells). Furthermore, as $d$ increases, $p_g$ is suppressed rapidly while $\hat{p}_g$ does not change much. When $d=7$, most individual window mispredictions do not cause global logical errors, but instead are paired up with each other and eliminated in the XOR operation.

We postulate that these mispredictions are caused by degeneracy at the seam between windows as discussed in Section~\ref{sec:without_merging}. When two decoding windows handle a syndrome configuration in a way that is different from the ground truth, but topologically equivalent to the ground truth and \emph{consistent} with each other, both windows would mispredict their own label $y_i$, but the combined global prediction would remain correct. This happens more often for the bulk window since it has two seams instead of one, as confirmed by our data.

This also indicates that the neural network does not learn the errors of individual windows completely independently, but rather develops a certain self-coordinating decoding among them. The effectiveness of self-coordination is further discussed later in Section.~\ref{sec:ablation_study}.

\subsection{Improvement of fault-tolerant threshold}
\begin{figure}[htbp]
\begin{center}
\includegraphics[width=0.4\linewidth]{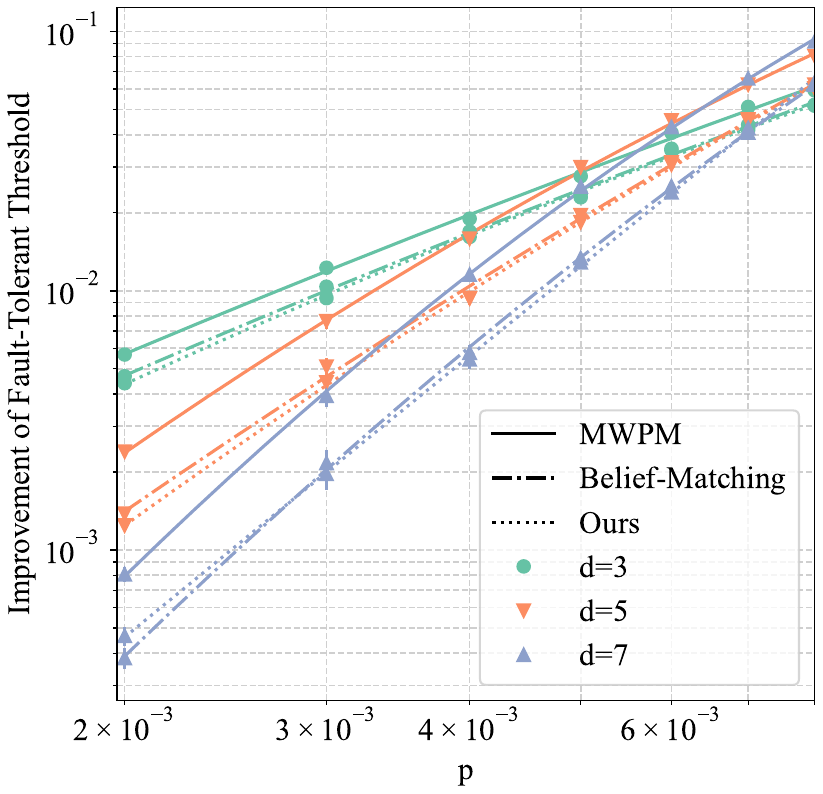}
\caption{Threshold improvement of our neural network decoder.}
\label{fig:threshold_all}
\end{center}
\end{figure}

We conduct the standard memory experiment to demonstrate our neural decoder's improvement of quantum error correction threshold. Since our model is designed for parallel window decoding and trained for at least 3 decoding windows, we still set the memory rounds as $N = 3d$ (splitting into 3 decoding windows to our neural decoder) and test the logical error rates. After that, we normalize them as the logical error rates for $d$ rounds of memory experiment to align with Figure~\ref{fig:threshold}. As shown in Figure~\ref{fig:threshold_all}, our decoder achieves or even surpasses the fault-tolerance threshold of Belief-Matching. Compared with MWPM, the threshold increases from approximately $\mathbf{0.6\%}$ to $\mathbf{0.7\%}$. This indicates that, at the current hardware error noise level~\cite{google2025quantum, gao2025establishing}, using a neural network decoder still holds significant value for experimental demonstrations.

\subsection{Temporal scalability}
\label{sec:arbitrarily_long_memory_experiment}
Different from AlphaQubit's serial pipeline, our decoder utilizes the parallel decoding scheme to generalize to arbitrary syndrome measurement rounds: Each decoding thread only decodes a single decoding window in a streaming manner, the 1-bit logical results returned by different threads are simply gathered and XORed together in the end to get the final logical correction. 

We test the temporal scalability of our decoder under this parallel decoding scheme by increasing the number of rounds in memory experiments until the logical error rate becomes close to $0.5$. We model the dependence of the final logical error rate $p_L$ on the number of rounds $N$ in the same way as~\cite{bausch2024learning}: We assume that this dependence can be approximately characterized with a single parameter, the logical error rate per round (LER) $\epsilon$, such that every round of memory experiment has a probability $\epsilon$ to \emph{flip} the result. The relationship between $p_L$, $\epsilon$, and $N$ is then given by:
\begin{equation}
    p_{L} =\frac{1 - (1-2\epsilon)^N}{2}, \epsilon = \frac{1}{2} (1 - \sqrt[N]{1 - 2 p_L}).
\end{equation}
Note that this implies that $p_L$ is always less than $0.5$. We assume that values of $p_L \ge 0.5$ observed in experiments are due to statistical fluctuation. Thus following~\cite{bausch2024learning}, we also define the \emph{fidelity} $F$ as $1 - 2 p_L$.

\begin{figure}
\begin{center}
    \subfigure[Decoding accuracy of $d = 3$]{
        \includegraphics[width=0.95\linewidth]{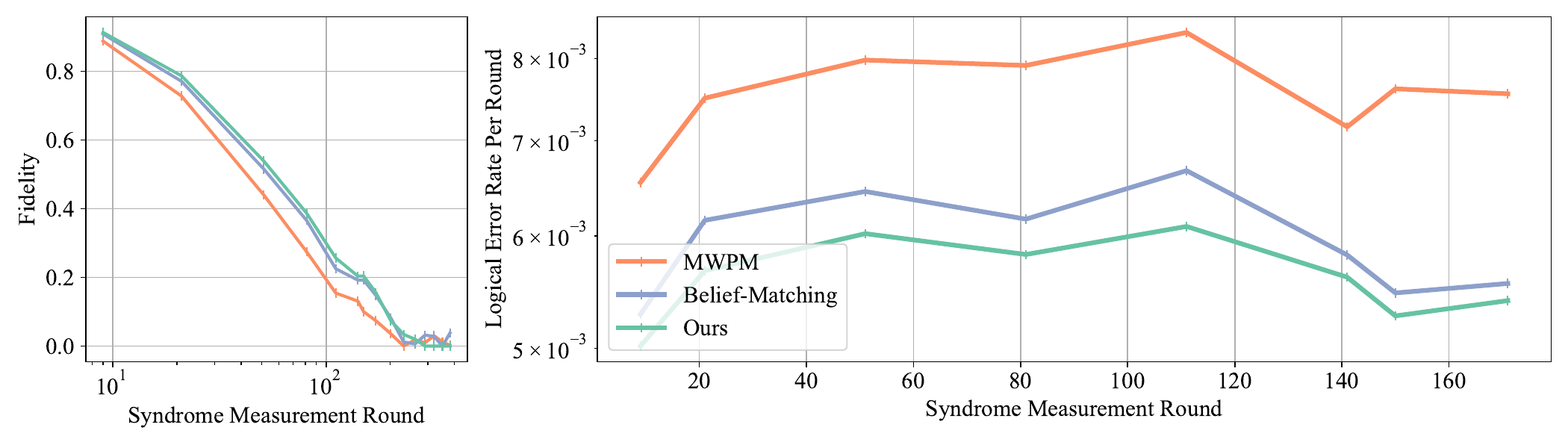}
        \label{fig:acc_d_3_effp}
    }
    \subfigure[Decoding accuracy of $d = 5$]{
        \includegraphics[width=0.95\linewidth]{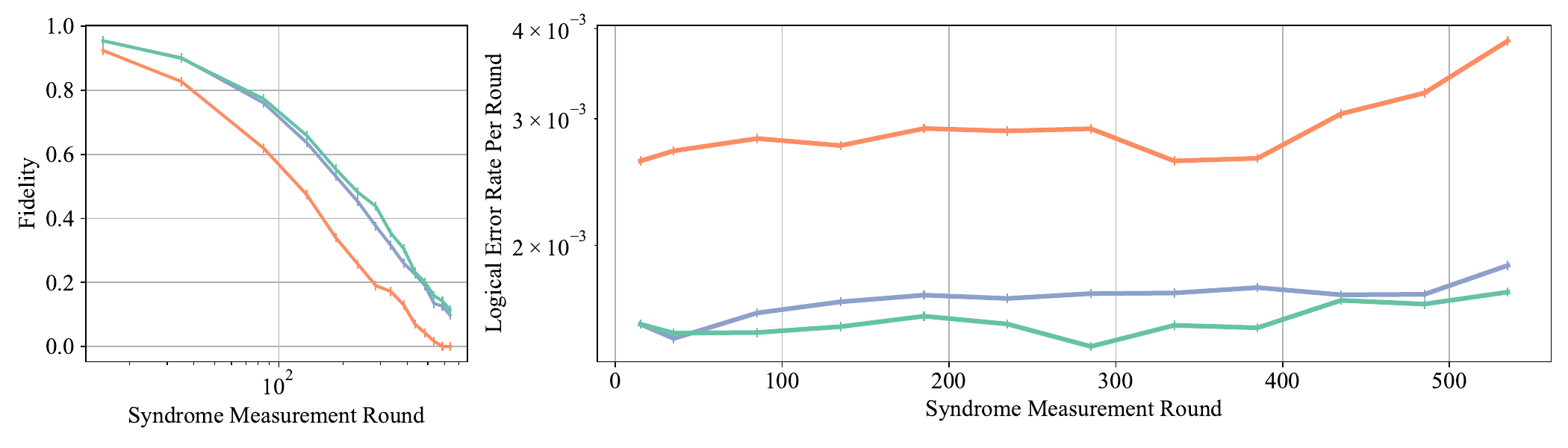}
        \label{fig:acc_d_5_effp}
    }
    \subfigure[Decoding accuracy of $d = 7$]{
        \includegraphics[width=0.95\linewidth]{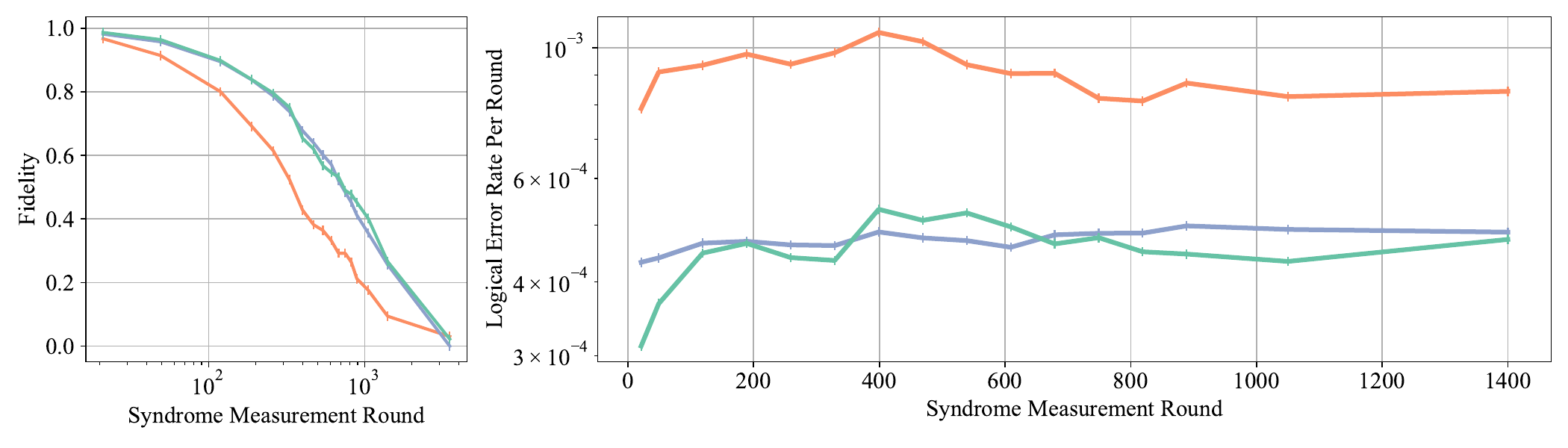}
        \label{fig:acc_d_7_effp}
    }
    \caption{Logical error rate per round for long time memory experiments under the effective physical error rates. We only plot the logical error rate per round for fidelity $F > 0.1$.}
    \label{fig:logical_error_rate}
\end{center}
\end{figure}

We plot the fidelity and the LER of our decoder, as well as two traditional global decoders, against the number of rounds in Figure~\ref{fig:logical_error_rate}. The LER remains roughly constant without any trend of degradation over time, suggesting that parallel decoding without local merging does not noticeably hurt the accuracy even as the number of decoding windows increases. Our decoder also significantly outperforms pure MWPM, and slightly outperforms Belief-Matching, which is about the same level as the reported performance of AlphaQubit~\cite{bausch2024learning}. We believe that it is fair to claim that our decoding scheme (which can be easily parallelized, unlike AlphaQubit) maintains near state-of-the-art accuracy when generalizing to arbitrary syndrome measurement rounds.

\subsection{Ablation studies}
\label{sec:ablation_study}
While AlphaQubit~\cite{bausch2023learning, bausch2024learning} has provided a comprehensive collection of ablation studies of different modules in the model design, we are focusing on several critical components in our modifications and designs. Since training a model from scratch entails considerable computational resources and time costs, which grow almost exponentially with the code distance $d$, we perform the ablation studies only on the $d = 3$ model.

The following model variations are trained and evaluated:
\begin{itemize}
    \item \textbf{Baseline}: The baseline model corresponding to Figure~\ref{fig:neural_network_architecture} and Table~\ref{tab:model_parameter}.
    \item \textbf{PEARes}: Putting the positional encoding layer (Section~\ref{sec:positional_encoding}) ahead of the \emph{ResNet} module (Section~\ref{sec:conv1d}).
    \item \textbf{BNTRST}: In batch normalization layers, setting \texttt{track\_running\_stat} to True.
    \item \textbf{SepWindow}: Utilizing independent trained models rather than one model to predict different kinds of decoding windows, to validate the effectiveness of self-consistency.
    \item \textbf{NoMixTrain}: Training process is designed to train three kinds of window data in different stages separately, each with $1/3$ training samples.
    \item \textbf{NoRec}: Training without the recurrent training method (Section~\ref{sec:recurrent_training}).
\end{itemize}

\begin{figure}[htbp]
\begin{center}
\includegraphics[width=0.75\linewidth]{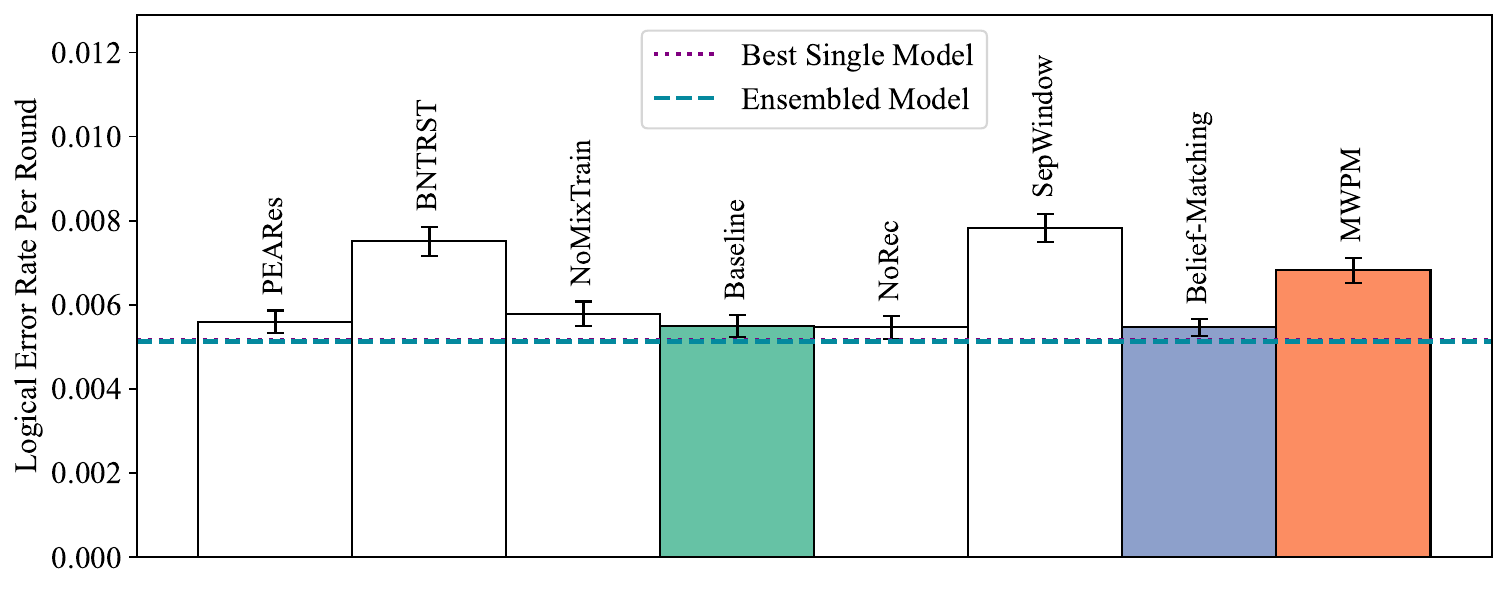}
\caption{Ablation study. Logical error rate per round of different model variations are tested, along with MWPM and Belief-Matching. The purple and blue dash lines are accuracy of the best single model (trained up to 40 GPU hours) and the ensembled model from Table~\ref{tab:ensemble}. }
\label{fig:ablation_study}
\end{center}
\end{figure}

To ensure a fair comparison, all models were trained across three datasets of $20$ million samples each (corresponding to $p = 0.006$, $p = 0.005$, and $p = 0.004$), with approximately $5$ GPU hours of training time allocated to each model.
Each variation's logical error rate per round under $p = p_\text{eff}(3)$ is illustrated in Figure~\ref{fig:ablation_study}. 

We summarize our main observations from the ablation study as follows:
\begin{enumerate}[label=(\arabic*)]
    \item The order between \textbf{PE} and \emph{ResNet} appears to be not as critical as we have imagined. 
    \item Disabling \texttt{track\_running\_stat} in BatchNorm is essential, likely because batches seen by the BatchNorm layer in training are not identically distributed due to the RNN architecture. A possible explanation is that our input representation requires the RNN to ``remember'' the current round number, which causes this information to affect the decoder state. Therefore, enabling \texttt{track\_running\_stat} makes the network behave too differently in evaluation from in training.
    \item If different models are trained completely independently for window decoding, then even though each single model has converged, the global accuracy after parallel decoding remains low due to the lack of self-coordination, or self-consistency. 
    \item Staging the training by window types or using the mixed samples directly has little influence on the final model capacity. Nevertheless, as discussed in Section~\ref{sec:individual_window} before, the training of bulk windows deserves more attention. 
    \item Although recurrent training is not so critical when the code distance is small, at $d=7$, we observe that models trained with multi-layer recurrent prediction tend to converge much faster, whereas models trained directly in a singular prediction often fail to converge as the initial training landscape is too complex. 
\end{enumerate}

\subsection{Model ensembling}

\begin{table}[h!]
\centering
\caption{Model ensembling and logical error rate per round.}
\label{tab:ensemble}
\begin{tabular}{ccccc}
\toprule
$d$ & Pymatching & Belief Matching & Best Single NN & Ensembled NN \\
\midrule
3 & 0.006260 & 0.005479 & 0.005152 & \textbf{0.005128} \\
5 & 0.002470 & 0.001442 & 0.001290 & \textbf{0.001276} \\
7 & 0.000912 & 0.000447 & 0.000437 & \textbf{0.000423} \\
\bottomrule
\end{tabular}
\end{table}

We also test the model ensembling method to improve the accuracy of logical prediction in Table~\ref{tab:ensemble}, where the ensembled model achieves the lowest logical error rate. The ensembling strategy is quite straight forward as we just average the output of top-$K$ ($K$ is chosen to be $5$ in our test, more is better) single models for each code distance $d$. We perform binary classification on the averaged output using $0.5$ as the threshold, the same during the single model test. More advanced ensembling strategies such as Bagging~\cite{breiman1996bagging, breiman2001random} and Boosting~\cite{freund1997decision} could further improve the logical accuracy.

\section{Discussion}
\label{sec:discussion}
In this study, we put forward a practical solution to address the throughput challenge of AlphaQubit, with insights drawn from parallel decoding schemes. Without modifying AlphaQubit's network architecture, our approach retains the model's key advantage of high accuracy while enabling parallelized, real-time decoding for neural network decoders---in turn enabling fault-tolerant quantum computing using AlphaQubit-type decoding systems.

Our work presents not only the first parallel approach to enabling arbitrary long quantum memory with AlphaQubit-type decoders, but also the first feasible solution for applying these decoders to logical operations. As the lattice surgery~\cite{litinski2019game} decoding graph is also modular and can be separated into some limited types of windows~\cite{bombin2023modular, tan2024sat}, we can train a neural network decoder for different types of decoding windows with sufficient buffer size $\geq d$. Afterwards, the decoding results from each window can be gathered to get the logical feedback. As long as the decoding process can be processed in a streaming and parallel pipeline, the reaction time~\cite{google2025quantum, wu2023fusion} can always be made constant regardless of the temporal-spatial volume of the decoding graph.

In this work, we conduct training up to a code distance of $d=7$ to demonstrate the feasibility of our approach. As the code distance scales and physical error rates further decrease, the required data volume, computational resources, and training time will all grow rapidly. Compared to the original AlphaQubit---which demonstrated training up to $d=11$---our network architecture does not present any additional fundamental barriers to training at that scale, and likely even for a few sizes larger, sufficient for early fault-tolerant quantum computing. Rather, we believe the challenges involved are simply a matter of resource allocation and computational speed. However, to advance to a code distance of $d=25$ (and beyond, when accounting for lattice surgery) in order to achieve the desired accuracy, further improvements in network design and sample efficiency must be considered. These enhancements are critical to fully integrating sophisticated neural networks with quantum devices.

Finally, although our network architecture is similar to AlphaQubit, the supervision for each window must be generated from ground truth logical errors in \emph{intermediate} syndrome measurement rounds, which are only available in simulated experiments, as opposed to \emph{final} logical error information available in real hardware experiments. 
Therefore, our current implementation cannot yet support end-to-end fine-tuning on experimental data in the absence of a calibrated circuit-level noise model, and we leave this for future work.

\section*{Acknowledgements}
The work is supported by National Key Research and Development
Program of China (Grant No.\ 2023YFA1009403), National Natural Science
Foundation of China (Grant No.\ 12347104), Beijing Natural Science Foundation
(Grant No.\ Z220002), Zhongguancun Laboratory, and Tsinghua University.
K.Z. would like to thank Haoran Wang for his valuable early-stage suggestions on the neural network training.

\bibliographystyle{quantum}
\bibliography{reference}

\onecolumn\newpage
\appendix

\section{Proof sketch for the fault tolerance of sandwich without merge}
\label{sec:no_merge}
\subsection{Preliminaries}

In the context of a QEC procedure, we call an event $A$ \emph{exponentially unlikely} if there exists a threshold $p_\text{th} > 0$, such that
$$\Pr(A) = O\left(\text{poly}(d)\cdot \left(\frac{p}{p_\text{th}}\right)^{\lceil d/2\rceil}\right).$$

Given a decoding graph $G = (V, E)$ and a subset $C \subseteq E$, we define the \emph{boundary} of $C$ as $\partial C = \{ v\in V \mid |\{e \in C \mid v \in e\}| \bmod 2 = 1\}$. We adopt the (slightly unusual but well-motivated) convention that ``virtual vertices'' of the entire decoding graph are not in $V$ (one can consider $G$ as a hypergraph with some edges with size $1$), and thus not in any $\partial C$, but ``virtual vertices'' on the time boundary of a window are in $V$ and thus can be in $\partial C$.

\medskip

We note that sandwich decoder~\cite{Tan2022ScalableSD} without merge will probably \emph{not} work if the inner decoder is allowed to connect the same two detection events across the seam with different paths in two adjacent windows. Therefore, for the following proof, we need an assumption about the inner decoder:

\begin{assumption}
\label{assumption:consistency}
    Let $G_1 = (V_1, E_1)$ and $G_2 = (V_2, E_2)$ be the decoding graphs of two adjacent windows with overlap, and let $C_1 \subseteq E_1$ and $C_2 \subseteq E_2$ be the corrections output by the inner decoder in these windows. For any $C'_1 \subseteq C_1$ and $C'_2 \subseteq C_2$, If $\partial C'_1 = \partial C'_2$, then $C'_1 = C'_2$. 
\end{assumption}

In plain words, if two subsets of corrections correct the same subset of detection events, then they are the same set of corrections. This property should hold for MWPM decoders if edge weights are perturbed to break ties.

\subsection{Proof sketch}
\begin{theorem}
\label{thm:min_weight}
    When Assumption~\ref{assumption:consistency} holds, any physical error configuration that causes the sandwich decoder with the MWPM inner decoder produces any non-trivial seam syndrome must have at least total weight $w_b/2$, where $w_b$ is the weighted buffer size (i.e., the shortest weighted distance on a window's decoding graph from a seam vertex to a virtual time boundary vertex).
\end{theorem}

Note that on a realistic decoding graph of the memory experiment, the weighted buffer size should be the buffer size times the weight of a vertical edge.

\begin{proof}[Proof sketch]
Consider two adjacent windows $G_1 = (V_1, E_1)$ and $G_2 = (V_2, E_2)$ with window corrections $C_1$ and $C_2$ respectively. Consider the symmetric difference $D = C_1 \oplus C_2$. Since $\partial C_1$ and $\partial C_2$ must agree in the region where the two windows overlap (where they much both agree with the actual detection events observed), it follows that $\partial D = \partial C_1 \oplus \partial C_2$ must be entirely outside of this overlap region.

Suppose that the sandwich decoder did produce some non-trivial seam syndrome, and let $v$ be any one detection event in the seam syndrome. Then $D$ must touch $v$. Let $D'$ be the connected component of $D$ that includes $v$. Let $C'_1 = D' \cap C_1$ and $C'_2 = D' \cap C_2$. Obviously $C'_1 \ne C'_2$, so by Assumption~\ref{assumption:consistency}, we must have $\partial C'_1 \ne \partial C'_2$, i.e., $\partial D' \ne \emptyset$. However, as mentioned above, $\partial D' \subseteq \partial D$ must entirely fall outside the overlap region. Take any $u \in \partial D'$. The distance between $v$ (on the seam) and $u$ (outside the overlap region) must be at least the weighted buffer size, $w_b$.

We have proven that there must exist a path in $D$ between $v$ and $u$. The path has at least length $w_b$ and is composed of edges in $C_1$ and $C_2$, so either $w_{C_1} \ge w_b/2$ or $w_{C_2} \ge w_b/2$. Since the inner decoder is MWPM, the total weight of any correction must be no more than the total weight of the physical error configuration, so the latter must also be at least $w_b/2$.
\end{proof}

\subsection{Discussion}
We note that Theorem~\ref{thm:min_weight} may not be very useful in practice, since with constant physical error rate and $O(d^3)$ possible error locations, there will almost certainly be much more than $b$ physical errors in any window. What we really want to prove is:
\begin{conjecture}
\label{conj:fault_tolerance}
    When the buffer size is at least $d$, and Assumption~\ref{assumption:consistency} holds, it is exponentially unlikely that the sandwich decoder with the MWPM inner decoder produces any non-trivial seam syndrome.
\end{conjecture}

Ideally, we want to be able to use a counting argument like the one in~\cite{fowler2012proof}. However, this would require characterizing the locations of the actual physical errors (e.g., at least $m/2$ errors on a length-$m$ path), not just their total weight. An apparent difficulty is that, while the path in $D$ must reach $u$ from $v$ without touching a space boundary (if the path touches a space boundary on \emph{both} sides before exiting the overlap region, then both windows should decode the path in the same way), the actual physical errors can touch the space boundary. Even though the total weight of physical errors still must be at least equal to the total weight of corrections, there seems to be more freedom with the path it takes.

\subsection{Idea to prove Conjecture~\ref{conj:fault_tolerance}}
\begin{proof}[Proof sketch]
We follow the proof of Theorem~\ref{thm:min_weight} up to the point where we get a path in $D$ between $v$ and $u$. Without loss of generality, we assume that $u$ is the only vertex on the path that falls outside of the overlap region. We denote that path as $P$, and its length (total weight) as $w_P$. We further denote $C_i \cap P$ as $C^P_i$. Then either $w_{C^P_1} \ge w_P/2$ or $w_{C^P_2} \ge w_P/2$. Without loss of generality, suppose that the former holds.

Below, we redefine the symbol $\partial$ so that it only includes \emph{real vertices} in the window $G_1$. This ensures that $\partial \mathcal{E} = \partial C_1$.

Now we find a subset $\mathcal{E}^Q$ of the physical errors $\mathcal{E}$ such that $\partial \mathcal{E}^Q = \partial C^Q_1$, for some $C^Q_1$ satisfying $C^P_1 \subseteq C^Q_1 \subseteq C_1$. We construct $\mathcal{E}^Q$ and $C^Q_1$ iteratively, starting from $\mathcal{E}^Q = \emptyset$ and $C^Q_1 = C^P_1$. Every step, as long as $\partial \mathcal{E}^Q \ne \partial C^Q_1$, we choose a vertex $t \in \partial \mathcal{E}^Q \oplus \partial C^Q_1$ arbitrarily (depending only on the current values of $\mathcal{E}^Q$ and $C^Q_1$). Since $\partial \mathcal{E} = \partial C_1$, we have that $t \in \partial (E - \mathcal{E}^Q) \oplus \partial (C_1 - C^Q_1)$. Therefore we can arbitrarily choose an edge incident to $t$ in either $E - \mathcal{E}^Q$ or $C_1 - C^Q_1$, and add it to $\mathcal{E}^Q$ or $C^Q_1$ respectively. This procedure must end because there are finitely many edges in $\mathcal{E}$ and $C_1$, and must result in $\partial \mathcal{E}^Q = \partial C^Q_1$.

Now observe that, \emph{a priori} (i.e., with $\mathcal{E}$, $\partial \mathcal{E}$, $C_i$... all unknown), all possible pairs $(\mathcal{E}^Q, C^Q_1)$ can be generated with a two-phase procedure.
\begin{itemize}
    \item First choose an arbitrary vertex $v$. Starting at $v$, every step move to an adjacent vertex, and choose whether the edge just traversed belongs to $C^P_1$ or $C^P_2$. Repeat this step until we reach a vertex $u$ outside of the overlap region, and we have the path $P$ as well as $C^P_1$.
    \item Then do the construction procedure in the previous paragraph, except that every step after determining $t$, we arbitrarily choose an edge incident to $t$ and arbitrarily choose either it belongs to $\mathcal{E}^Q$ or $C^Q_1$. When this procedure ends, we get a candidate $(\mathcal{E}^Q, C^Q_1)$.
\end{itemize}

Consider all such candidates where the whole procedure takes $n$ steps. Letting $w_\text{min}$ be the minimum weight of any edge in the decoding graph, and $n_1$ and $n_2$ be the number of steps each phase takes (thus $n_1 + n_2 = n$), we have:
$$w_P \ge n_1 \cdot w_\text{min}, \qquad w_{C^P_1} \ge \frac{n_1}{2} \cdot w_\text{min},$$
$$w_{\mathcal{E}^Q} + w_{C^Q_1} \ge (\frac{n_1}{2} + n_2) \cdot w_\text{min}, \qquad w_{\mathcal{E}^Q} \ge (\frac{n_1}{4} + \frac{n_2}{2}) \cdot w_\text{min} \ge \frac{n}{4} \cdot w_\text{min},$$
where the last step makes use of the fact that $C^Q_1$ is a min-weight correction for $\partial \mathcal{E}^Q = \partial C^Q_1$. The probability that all edges in the candidate $\mathcal{E}^Q$ are in the actual $\mathcal{E}$ is upper bounded by $\hat{p}_n = \exp \left(\frac{n}{4} \cdot w_\text{min}\right)$. Meanwhile, letting $k$ be the maximum degree of any vertex in the decoding graph, the number of all candidates is upper bounded by $\hat{N}_n = |V_\text{seam}| \cdot (2k)^{n}$. The total probability of those candidates are thus lower bounded by
$$\hat{p}_n \cdot \hat{N}_n = |V_\text{seam}| \cdot \left(2k\exp \frac{w_\text{min}}{4}\right) ^ n.$$

Therefore $\hat{p}_n \cdot \hat{N}_n$ is a geometric series, and for fixed $k$, when $w_\text{min}$ is large enough (i.e., when the physical error rate $p$ of the most likely error mechanism is low enough), this series decays exponentially with $n$. Furthermore, $n$ is lower bounded by the (unweighted) buffer size $b$\footnote{More accurately, $n_1$ is lower bounded by $b$ and $n_2$ is lower bounded by $\frac{w_b}{2w_\text{min}}$.}. By taking $b = 4\lceil d/2\rceil$, the leading term of this geometric series has the form
$$\hat{p}_b \cdot \hat{N}_b = |V_\text{seam}| \cdot \left((2k)^4\exp w_\text{min}\right) ^ {\lceil d/2\rceil} = O\left(\text{poly}(d)\cdot O(p)^{\lceil d/2\rceil}\right).$$
We can let $p$ be small enough that the common ratio of the geometric series is upper bounded by a constant (e.g., 1/2), so that the sum of this geometric series is still an exponentially unlikely probability.

\end{proof}

\section{Training scaling law with respect to code distance and testing physical error rate}
\label{sec:train_log}

We recorded and illustrated the testing logical error rates per round during training and the corresponding GPU hours for $d = 3, 5, 7$, as shown in Figure~\ref{fig:training_curve} below.

\begin{figure}
\begin{center}
    \subfigure[Training log of $d = 3$]{
        \includegraphics[width=0.98\linewidth]{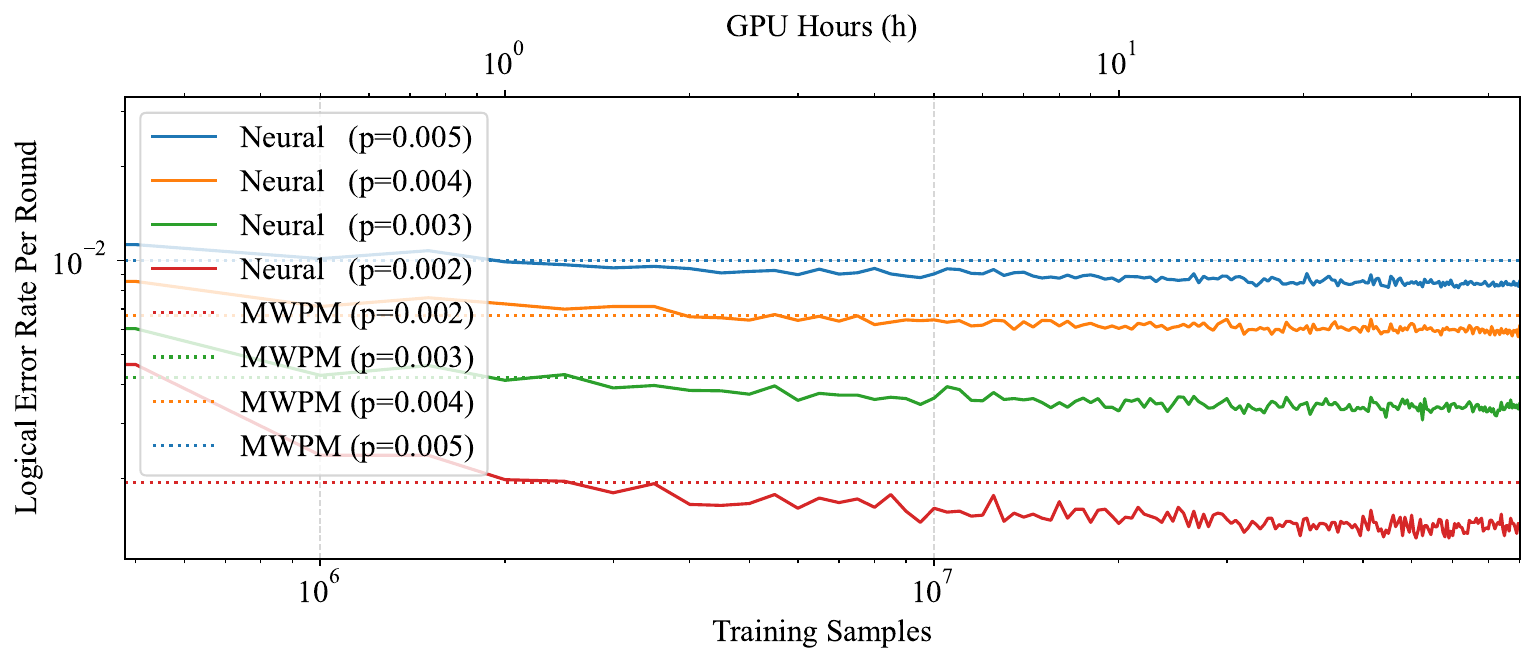}
    }
    \subfigure[Training log of $d = 5$]{
        \includegraphics[width=0.98\linewidth]{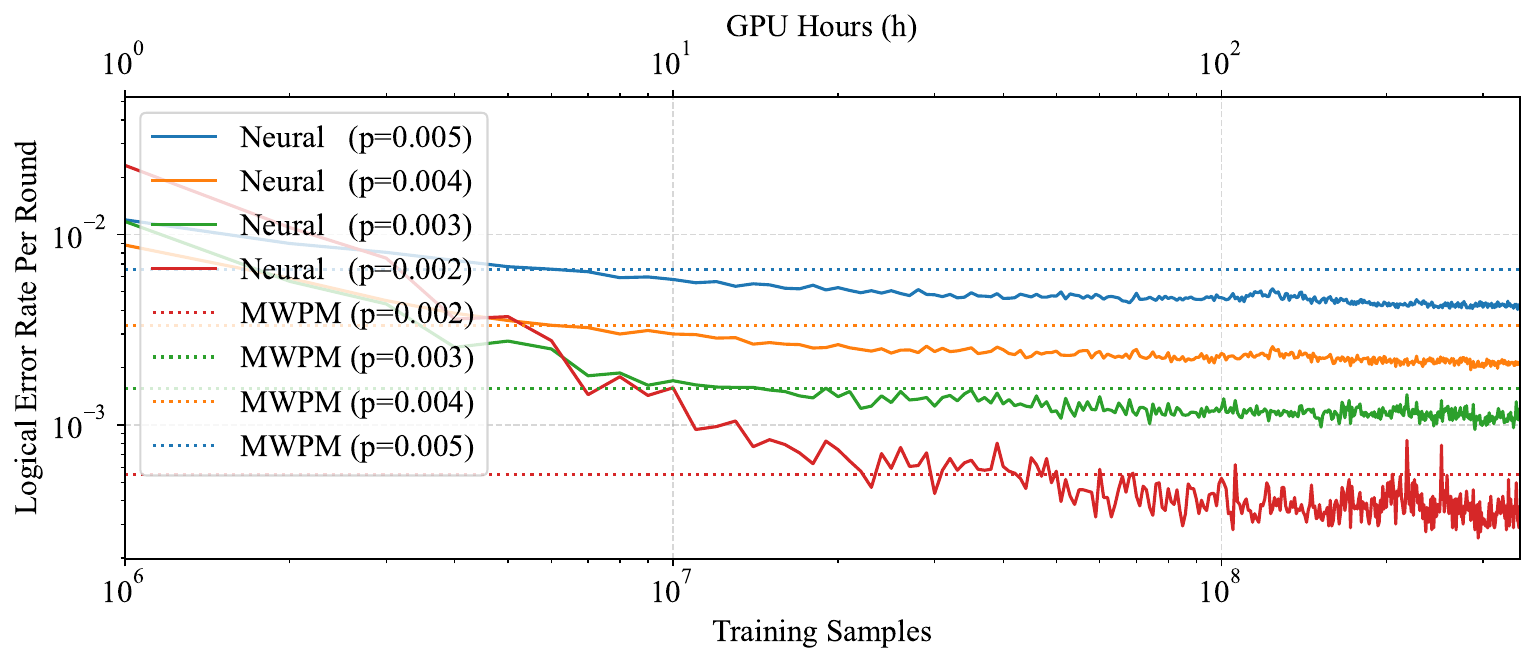}
    }
    \subfigure[Training log of $d = 7$]{
        \includegraphics[width=0.98\linewidth]{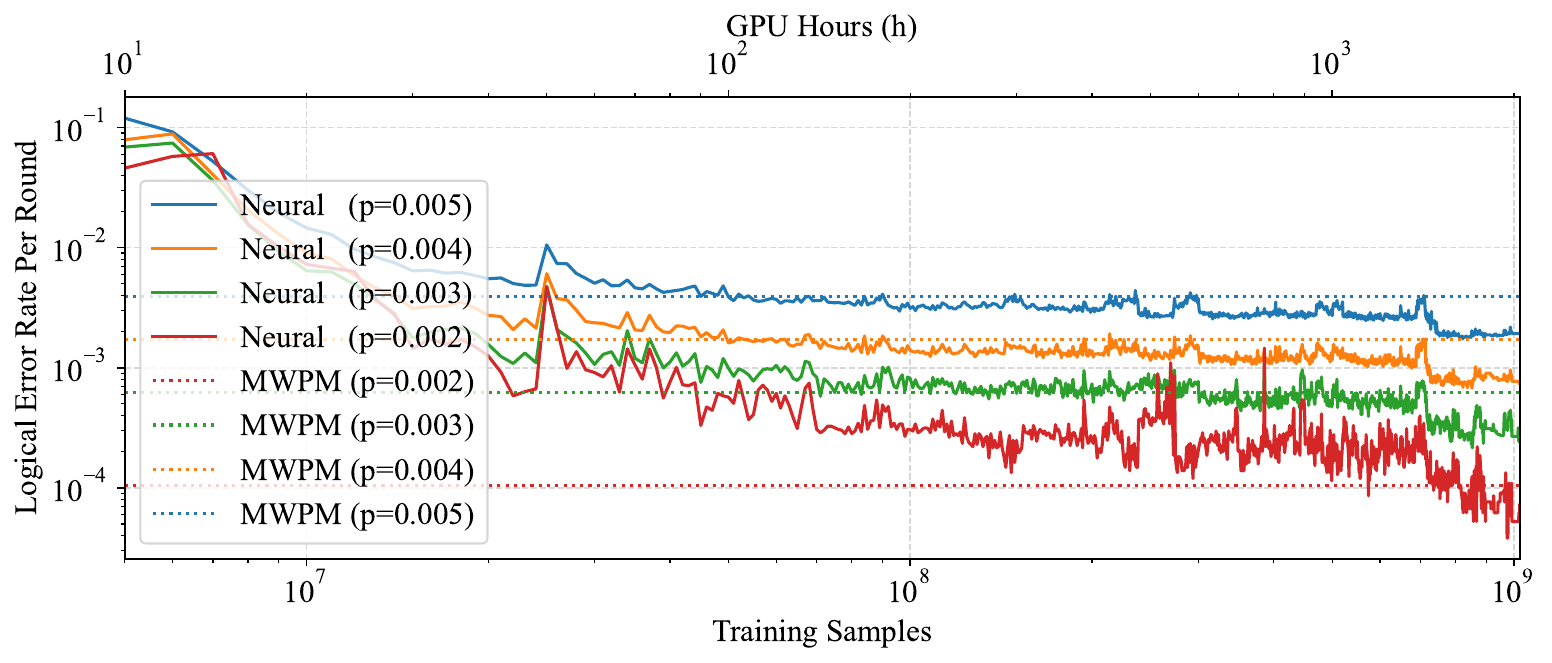}
    }
    \caption{Detailed training logs for code distance $d=3,5,7$ and testing logical error rates for physical error rate $p = \{0.002, 0.003, 0.004, 0.005\}$. As $d$ increases or $p$ decreases, the required number of training samples to match the performance of classical decoding algorithms increases exponentially.}
    \label{fig:training_curve}
\end{center}
\end{figure}

\end{document}